\documentclass[nonacm,acmsmall]{acmart}

\AtBeginDocument{%
  \providecommand\BibTeX{{%
    \normalfont B\kern-0.5em{\scshape i\kern-0.25em b}\kern-0.8em\TeX}}}

\setcopyright{none}
\acmPrice{}

\renewcommand\footnotetextcopyrightpermission[1]{} 
\makeatletter
\let\@authorsaddresses\@empty
\makeatother

\usepackage{url}            %
\usepackage{nicefrac}       %

\usepackage{enumitem}
\usepackage[ruled, linesnumbered]{algorithm2e}
\usepackage{amsthm, amsmath, amsfonts, mathtools}
\usepackage{bbm}
\usepackage{hyperref}
\usepackage[capitalise,noabbrev,nameinlink]{cleveref}
\usepackage{xcolor}
\usepackage{chngcntr}
\usepackage{apptools}

\newtheorem{lemma}{Lemma}
\newtheorem{theorem}{Theorem}
\newtheorem{corollary}{Corollary}
\AtAppendix{\counterwithin{lemma}{section}}

\theoremstyle{definition}
\newtheorem{definition}{Definition}

\theoremstyle{remark}
\newtheorem{remark}{Remark}
\newtheorem{example}{Example}
\newtheorem{observation}{Observation}

\newcommand{\Ption}{\textsc{Partition}\xspace}

\newcommand{\BDR}{\textsc{BarterDR}\xspace}

\newcommand{\CCWalk}{\textsc{CCWalk}\xspace}
\newcommand{\CCWalkp}[1]{$\text{\textsc{CCWalk}}(#1)$\xspace}
\newcommand{\FCCC}{\textsc{FindCCC}\xspace}

\newcommand{\GKPS}{\textsc{GKPS}\xspace}
\newcommand{\GKPSDR}{\textsc{GKPS-DR}\xspace}
\newcommand{\cf}{f}
\newcommand{\cN}{\mathcal{N}}
\newcommand{\cB}{\mathcal{B}}

\newcommand{\Z}{\mathbb{Z}}

\newcommand{\E}{\mathbb{E}}
\newcommand{\R}{\mathbb{R}}
\DeclareMathOperator{\poly}{poly}
\newcommand{\BSV}{$\mathsf{BarterSV}$\xspace}
\newcommand{\VBM}{$\mathsf{ValBalMatching}$\xspace}
\newcommand{\BSVcaps}{$\mathsf{BarterSV\text{-}Caps}$\xspace}

\newcommand{\BSVLP}{$\mathsf{BarterSV\text{-}LP}$\xspace}
\newcommand{\BSVLPcap}{$\mathsf{BarterSV\text{-}LP\text{-}Caps}$\xspace}

\newcommand{\CCC}{$\langle s_i\leadsto t_i\rangle_{i \in [q]}$\xspace}
\newcommand{\mCCC}{\langle s_i\leadsto t_i\rangle_{i \in [q]}}
\newcommand{\ind}{\mathbbm{1}\xspace}
\newcommand{\Oh}{\mathcal{O}}
\newcommand{\Pc}{\mathcal{P}}
\newcommand{\seq}[1]{\left< #1 \right>}

\DeclareMathOperator{\OPTIP}{OPT_{IP}}
\DeclareMathOperator{\OPTLP}{OPT_{LP}}

\begin{document}

\title{Barter Exchange with Shared Item Valuations}

\author{Juan Luque}
\affiliation{%
  \department{Department of Computer Science}
  \institution{University of Maryland}
  \city{College Park}
  \state{MD}
  \country{USA}}
\author{Sharmila Duppala}
\affiliation{%
  \department{Department of Computer Science}
  \institution{University of Maryland}
  \city{College Park}
  \state{MD}
  \country{USA}}
\author{John Dickerson}
\affiliation{%
  \department{Department of Computer Science}
  \institution{University of Maryland}
  \city{College Park}
  \state{MD}
  \country{USA}}
\author{Aravind Srinivasan}
\affiliation{%
  \department{Department of Computer Science}
  \institution{University of Maryland}
  \city{College Park}
  \state{MD}
  \country{USA}}


\begin{abstract}
In barter exchanges agents enter seeking to swap their items for other items on their wishlist. We consider a centralized barter exchange with a set of agents and items where each item has a positive value. The goal is to compute a (re)allocation of items maximizing the agents' collective utility subject to each agent's total received value being comparable to their total given value. Many such centralized barter exchanges exist and serve crucial roles; e.g., kidney exchange programs, which are often formulated as variants of directed cycle packing. We show finding a reallocation where each agent's total given and total received values are equal is NP-hard. On the other hand, we develop a randomized algorithm that achieves optimal utility in expectation and where, i) for any agent, with probability 1 their received value is at least their given value minus $v^*$ where $v^*$ is said agent's most valuable owned and wished-for item, and ii) each agent's given and received values are equal in expectation. 
\end{abstract}

\maketitle
\thispagestyle{empty}

\section{Introduction}
Social media platforms have recently emerged into small scale business websites. For example, platforms like Facebook, Instagram, etc.  allow its users to buy and sell goods via verified business accounts. With the proliferation of such community marketplaces, there are growing communities for buying, selling and exchanging (swapping) goods amongst its users. 
We consider applications, viewed as Barter Exchanges, which allow users to exchange board games, digital goods, or any physical items amongst themselves. For instance, the subreddit \textsc{GameSwap}\footnote{\url{www.reddit.com/r/Gameswap}} (61,000 members) and Facebook group \textsc{BoardgameExchange}\footnote{\url{https://www.facebook.com/groups/boardgameexchange}} (51,000 members) are communities where users enter with a list of owned video games and board games. 
The existence of this community is testament to the fact that although users could simply liquidate their goods and subsequently purchase the desired goods, it is often preferable to directly swap for desired items. 
Additionally, some online video games have fleshed out economies allowing for the trade of in-game items between players while selling items for real-world money is explicitly illegal e.g., Runescape\footnote{www.jagex.com/en-GB/terms/rules-of-runescape}. In these applications, a centralized exchange would achieve greater utility, in collective exchanged value and convenience, as well as overcome legality obstacles.

A centralized barter exchange market provides a platform where agents can exchange items directly, without money/payments. Beyond the aforementioned applications, there exist a myriad of other markets facilitating the exchange of a wide variety of items, including books, children's items, cryptocurrency, and human organs such as kidneys. There are both centralized and decentralized exchange markets for various items. 
 \textsc{HomeExchange}\footnote{\url{www.homeexchange.com}}
and \textsc{ReadItSwapIt}\footnote{\url{www.readitswapit.co.uk}} are decentralized marketplaces that facilitate pairwise exchanges by mutual
agreement of vacation homes and books, respectively. Atomic cross chain swaps allow users to exchange currencies within or across various cryptocurrencies~\citep[e.g.,][]{Herlihy18:Atomic,Thyagarajan22:Universal}. Kidney exchange markets \citep[see, e.g.,][]{azizFairAllocationIndivisible2021,abrahamClearingAlgorithmsBarter2007} and children's items markets (e.g., \textsc{Swap}\footnote{\url{www.swap.com}}) are examples of centralized exchanges facilitating swaps amongst incompatible patient-donor pairs and children items or services amongst parents.  
Finding optimal allocations is often NP-hard. As a result heuristic solutions
have been explored extensively \cite{glorie2014kidney,plaut2016fast}.

Currently, the aforementioned communities \textsc{GameSwap} and \textsc{BoardGameExchange} make swaps in a decentralized manner between \textit{pairs} of agents, but finding such pairwise swaps is often inefficient and ineffective due to demanding a ``double coincidence of wants''~\citep{Jevons79:Theory}. However, centralized \textit{multi-agent} exchanges can help overcome such challenges by allowing each user to give and receive items from possibly different users.
Moreover, the user's goal is to swap a subset of their owned games for a subset of their desired games with comparable (or greater) value. 
Although an item's value is subjective, a natural proxy is its re-sale price, which is easily obtained from marketplaces such as Ebay.

We consider a centralized
exchange problem where each agent has a have-list and a wishlist of distinct (indivisible) items (e.g., physical games) and, more generally, each item has a value agreed upon by the participating agents (e.g., members of the \textsc{GameSwap} community). The goal is to find an allocation/exchange that (i) maximizes the collective utility of the allocation such that (ii) the total value of each agent's items before and after the exchange is equal.\footnote{Equivalently, the total value of the items given is equal to the total value of the items received.}
We call this problem \emph{barter with shared valuations}, \BSV, and it is our subject of study.  Notice that bipartite perfect matching is a special case of \BSV where each agent has a single item in both its have-list and wishlist each and where all the items are have the same value. On the other hand, we show \BSV is NP-Hard (\cref{thm:bsv-hardness}). %

In the following sections we formulate \BSV as bipartite graph-matching problem with additional \textit{barter constraints}. Our algorithm \BDR is based on rounding the fractional allocation of the natural LP relaxation to get a feasible integral allocation. 
A direct application of existing rounding algorithms (like \cite{gandhiDependentRoundingIts2006}) to \BSV results in a worst-case where some agents give away all their items and receive none in exchange. This is wholly unacceptable for any deployed centralized exchange. In contrast, our main result ensures \BDR allocations have reasonable net value for all agents; more precisely each agent gives and receives the same value in expectation and the absolute difference between given and received values is at most the value of their most valuable item (\cref{thm:bdr}).

\subsection{Problem formulation: \texorpdfstring{\BSV}{BarterSV}} \label{sec:prob-form}
Suppose we are given a ground set of items $\mathcal{I}$ to be swapped, item values $\{ v_j \in \R^+ : j \in \mathcal{I} \}$ where $\R^+$ denotes the non-negative real numbers, and a community of agents $\cN= [n]$ where $[n]$ denotes $\{1,2,\dots,n\}$. Each agent $i$ possesses items $H_i \subseteq \mathcal{I}$ and has wishlist $W_i \subseteq \mathcal{I}$.  
A \emph{valid allocation} of these items involves agents swapping their items with other agents that desire said item. Let $w(i,i',j)$ denote the utility gained when agent $i$ gives an item $j \in H_i \cap W_{i'}$ to an agent $i'$.
The goal of \BSV is to find a valid allocation of maximum utility \textit{subject to} no agent giving away more value than they received. Formally, any valid allocation is represented by the function $g: \cN \times \cN \times \mathcal{I} \rightarrow \{0,1\}$ such that $g(i,i',j)=1$ if \textit{agent $i$ gives agent $i'$ item $j$} and $g(i,i',j)=0$, otherwise.
Additionally, each agent $i \in \cN$  (i) receives only desired items, meaning $\sum_{i'\in \cN}g(i',i,j) \leq 1$ if $j \in W_{i}$, and $\sum_{i'\in \cN}g(i',i,j) \leq 0$ otherwise; and (ii) gives away only items they possess, thus $\sum_{i' \in \cN}g(i,i',j) \leq 1$ if $j \in H_i$, and $\sum_{i' \in \cN}g(i,i',j) \leq 0$ otherwise. 
We begin by defining a special variant of a graph matching problem called \textit{Value-balanced Matching (VBM)}. 
\begin{definition}{\textit{Value-balanced Matching (VBM)}} \label{def:vpm}
    Suppose there is bipartite graph $G = (L,R,E)$ with vertex values $v_a>0$ $\forall a \in L \cup R$. Let each edge $e\in E$ have weight $w_e \in \R$, $L =  \dot\bigcup_{i}L_i$, and $R = \dot\bigcup_{i}R_i $. For a given matching $M \subseteq E$, let ${V_i}^{(L)} =  \sum_{ \ell : (\ell,r) \in M, \ell \in L_i} v_\ell $ and ${V_i}^{(R)} =  \sum_{ r : (\ell,r) \in M, r \in R_i} v_r$,
    where $\dot\bigcup$ denotes disjoint union. The goal of VBM is to find $M$ of maximum weight subject to, for each $i$, the value of items matched in $L _i$ and $R_i$ are equal i.e., ${V_i}^{(R)} = {V_i}^{(L)}$. 
\end{definition}
\begin{figure} 
    \centering
    \includegraphics[width=.36\textwidth]{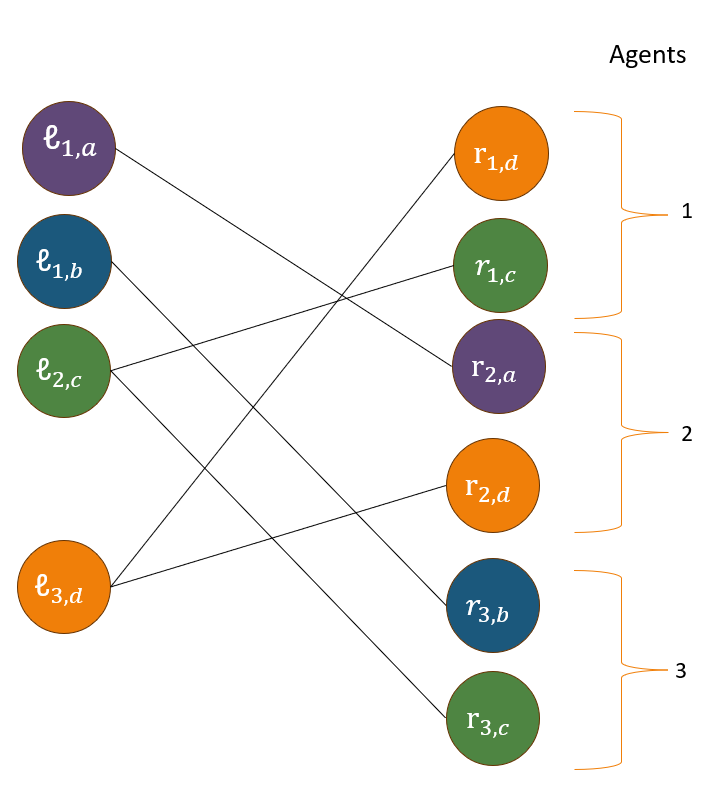}
    \caption{A VBM instance for a \BSV instance with $\mathcal{N} = \{1, 2\}$, $\mathcal{I} = \{ a, b, c, d \}$, $H_1 = \{ a, b\}$, $W_1 = \{ c, d\}$, $H_2 = \{ c\}$, $W_2 = \{ a, d \}$, $H_3 = \{ d\}$, $W_3 = \{b, c\}$, $v_a = 100 $, $v_b = v_c = v_d = 1$,
    and $w_e=1$ for all $e \in E$}
    \label{fig:vbm}
\end{figure}

\begin{lemma} \label{lemma:instance_equiv}
    For any instance $\mathcal{B}$ of \BSV, there exists a corresponding instance $\mathcal{G}$ of VBM such that the utility of an optimal allocation in the instance $\mathcal{B}$ is equal to the optimal weight matching in the corresponding instance $\mathcal{G}$ and vice-versa.
 \end{lemma}

\begin{proof}
   Given a \BSV instance $\mathcal{B}$ given by $(\cN,\mathcal{I}, \{v_j\}_{j\in \mathcal{I}},(H_i,W_i)_{i\in [n]},w)$ we can reduce it to a corresponding instance $\mathcal{G}$ of VBM via the construction of an appropriate bipartite graph with vertex values. For each agent $i \in \mathcal{N}$, build the vertex sets
$L_i := \{ \ell_{ij} : j \in H_i\}$ and 
$R_i := \{ r_{ij} : j \in W_i\}$. Then the bipartite
graph of interest has vertex sets
$L = \dot\bigcup_{i \in [n]} L_i $ and $R = \dot\bigcup_{i\in [n]} R_i$,
as well as edge set
$
E := \{ (\ell_{ij}, r_{i'j}) : j \in H_i \cap W_{i'} \wedge i, i' \in \mathcal{I}\} $.  For each edge $e = (\ell_{ij},r_{i'j})$, the weight of $e$ is given by $w_{e}:= w(i,i',j)$.
Each vertex $\ell_{ij}$ and $r_{i'j}$ has vertex values $v_j$ i.e., $v_{\ell_{ij}}, v_{r_{i'j}} := v_j $. Notice that the edges in \( E \) represent vertices belonging the same item, hence they have identical values. Refer to \cref{fig:vbm} for an example constructing a VBM instance. 

Suppose we are given any feasible allocation $g$ for an instance $\cB$ of \BSV with utility of $D$, then we can say construct a feasible matching $M$ corresponding to instance $\mathcal{G}$ of weight $D$. We construct the matching as follows. 
For any $i, i' \in \cN$ and $j\in H_i \cap W_{i'}$, if $g(i,i',j) =1$, then add the edge $e = (\ell_{ij},r_{i'j})$ to $M$.
The total weight of the matching is given by $\sum_{e \in M}w_e  = \sum_{ (\ell_{ij},r_{i'j}) \in M}w(i,i',j) = D$. 
It remains to show that $M$ is indeed a value-balanced matching. We begin by showing that $M$ is a feasible matching to the VBM problem. Since $g$ is a valid allocation to $\cB$, we can say that for any agent $i \in \cN$, the value of the items received is equal to the value of the items given away i.e., $\sum_{j \in H_i} \sum_{i' \in \cN }g(i,i',j) =\sum_{j \in W_i}\sum_{i' \in \cN}g(i',i,j) $. This implies that for each $g(i,i',j)=1$, we have $(l_{ij},r_{i'j}) \in M$, therefore, $V_i^{(L)} = V_{i}^{(R)}$. Thus, $M$ is a feasible VBM matching. 

To prove the other direction we begin by assuming that $M$ is a feasible matching to $\mathcal{G}$ of weight $D$. Given $M$, we can construct a feasible allocation $g$ by assigning $g(i,i',j)=1$ if the edge $(\ell_{ij},r_{i'j})\in M$, and $g(i,i',j)=0$ otherwise. Notice that $g$ is a valid allocation; for any agent $i \in \cN$, the value of items received is equal to the value of the items given away. This is because $M$ is a feasible matching to VBM, meaning for any $i\in [n]$, $V_i^{(L)} = V_{i}^{(R)}$ implying that $\sum_{i'\in \cN}\sum_{j \in H_i\cap W_{i'}}g(i,i',j) = \sum_{i' \in \cN} \sum_{j \in W_i\cap H_{i'}} g(i',i,j)$. 
The utility of allocation is given by $\sum_{i',i \in N}\sum_{j \in H_i\cap W_{i'}}w(i,i',j) = \sum_{e=(\ell_{ij},r_{i'j})\in M} w_e =D .$ This completes our proof. 
\end{proof}
\begin{corollary} \label{lemma:bsv-equiv-to-vbm}
    \BSV is equivalent to VBM.
\end{corollary} 
The proof of the corollary directly follows from \cref{lemma:instance_equiv}.

\subsection{LP formulation of VBM}
Any feasible matching $M$ in $(L,R,E)$ such that for each $e \in E$, $x_e = 1$ if $e \in M$ and $x_e=0$ otherwise is feasible in the following Integer Program (IP); 

\begin{subequations} \label{eq:barter-IP}
\begin{align}
    \max \quad            & \sum_{e \in E} w_e x_e \label{eq:LP-dynamic}                    \\
    \text{subj. to} \quad & x(\ell_{ij}) \le 1, & i\in [n], \ell_{ij} \in L_i&
    \label{eq:elli-caps}\\
                          & x(r_{ij}) \le 1, & i \in [n], r_{ij} \in R_i&
                          \label{eq:ri-caps}\\
                          & \sum_{a \in L_i} x(a) v_a = \sum_{b \in R_i} x(b) v_b, & i \in [n]& \label{eq:values-barter} \\
                          & x_e \in \{0,1\},
                          & e = (\ell_{ij}, r_{i'j})\in E&.
                          \label{eq:integrality-constr}
\end{align}
\end{subequations}
For $a \in L \cup R$, we denote $x(a) := \sum_{e \in N(a)} x_e$ where $N(a)$ denotes
the open neighborhood of $a$ i.e., $N(a) := \{ (a, b) \in E : b \in L \cup R\}$. 
Thus we can say that if an edge $e=(\ell_{ij}, r_{i'j}) \in M$ says agent $i$ gives item $j$ to agent $i'$ in the corresponding \BSV instance. 

\begin{lemma} \label{lemma:lp-is-ub}
IP \eqref{eq:barter-IP} is equivalent to \BSV.
Moreover, the objective of \BSVLP is an upper bound on the objective of IP~\eqref{eq:barter-IP}.
\end{lemma}
\begin{proof} 
IP~\eqref{eq:barter-IP} is clearly equivalent to VBM and VBM is equivalent to \BSV, per \cref{lemma:bsv-equiv-to-vbm}.
The objective of IP~\ref{eq:barter-IP}, $\sum_{e \in E} w_e x_e$ is the \emph{allocation's utility}. 
By relaxing \eqref{eq:integrality-constr} to $x_e \ge 0$ for $e \in E$ we arrive at the natural LP relaxation of \BSV, namely \BSVLP, which can only have a greater optimal objective value. 
\end{proof}

We conclude this section by a simple note. 
For each $e=(\ell_{ij},r_{i'j}) \in E$ we may set $w_{e} = v_j$ and recover the objective of maximizing the collective value received by all agents. Nevertheless, 
our results hold even if $w_e$ is set arbitrarily.
For example, the algorithm designer could place greater value on certain item transactions, or they may maximize the sheer number of items received by uniformly setting $w_e=1$.

\section{Preliminaries: GKPS dependent rounding}
\label{sec:prelim-gkps}
Our results build on the dependent rounding algorithm due to \cite{gandhiDependentRoundingIts2006}, henceforth referred to as \GKPSDR. \GKPSDR is an algorithm that takes $\{x_e\}\in [0,1]^{|E|}$ defined over the edge set $E$ of a biparite graph $(L, R, E)$ and outputs $\{X_e\} \in \{0,1\}^{|E|}$. 
In each iteration \GKPSDR considers the graph of floating edges (those edges $e$ with $0 < x_e < 1$) and selects a maximal path or cycle $P \subseteq E$ on floating edges. The edges of $P$ are decomposed into alternate matchings $M_1$ and $M_2$ and rounded
in the following way. Fix
$\alpha^{\text{GKPS}} = \min \left\{ \gamma > 0 : \left(\bigvee_{e\in M_1} x_e + \gamma = 1\right) \vee \left(\bigvee_{e \in M_2} x_e - \gamma = 0\right) \right\}$,
and 
$\beta^{\text{GKPS}} = \min \left\{ \gamma > 0 : \left(\bigvee_{e\in M_2} x_e + \gamma = 1\right) \vee \left(\bigvee_{e \in M_1} x_e - \gamma = 0\right) \right\}$.
Thus, each $x_e$ is updated to $x_e'$ according to one of the following disjoint events:
with probability $\frac{\beta^{\text{GKPS}}}{\alpha^{\text{GKPS}} + \beta^{\text{GKPS}}}$
\[
x_e' = 
\begin{cases}
    x_e + \alpha, &e \in M_1\\
    x_e - \alpha, &e \in M_2
\end{cases};
\quad\text{else,}
\quad
x_e' = 
\begin{cases}
    x_e - \beta, &e \in M_1\\
    x_e + \beta, &e \in M_2.
\end{cases}
\]
The selection of $\alpha$ and $\beta$ ensures at least one edge is rounded to $0$ or $1$ in every iteration. \GKPSDR guarantees \ref{prop:marginal} \emph{marginal}, \ref{prop:degree} \emph{degree preservation}, and \ref{prop:neg-corr} \emph{negative correlation} properties:
\begin{enumerate}[label=\textbf{(P\arabic*)}]
    \item $\forall e\in E$, $\Pr(X_e = 1) = x_e$. \label{prop:marginal}
    \item $\forall a \in L \cup R$ and with probability 1, $X(a) \in \{ \lfloor x(a) \rfloor, \lceil x(a) \rceil  \}$. \label{prop:degree}
    \item $\forall a \in L \cup R,\; \forall S \subseteq N(a),\; \forall c \in \{0,1\}$,
    $\Pr\left(\bigwedge_{s \in S} X_s = c\right) \le  \prod_{s \in S} \Pr\left( X_s = c\right)$. \label{prop:neg-corr}
\end{enumerate}
\begin{remark} \label{remark:gkps-path-length}
When \GKPSDR rounds a path between vertices $a$ and $b$, the signs of the changes to $x(a)$ and $x(b)$ are equal if and only if 
$a$ and $b$ belong to different graph sides.
\end{remark}

\section{Related work} \label{sec:related-work}
Centralized barter exchanges have been studied by several others in the context of kidney-exchanges \cite{abrahamClearingAlgorithmsBarter2007,ashlagiEffectMatchrunFrequencies2018,azizFairAllocationIndivisible2021}. 
\BSV generalizes a well-studied kidney-exchange problem in the following way. The Kidney Exchange Problem (KEP) is often formulated as directed cycle packing in compatibility patient-donor graphs \cite{abrahamClearingAlgorithmsBarter2007} where each node in the graph corresponds to a patient-donor pair and directed edges between nodes indicate compatibility. \citet{abrahamClearingAlgorithmsBarter2007,biroMaximumWeightCycle2009} observed this problem reduces to
bipartite perfect matching, which is solvable in polynomial-time.
We show \BSV is NP-Hard and thus resort to providing a randomized algorithm with approximate guarantees on the agents' net values via LP relaxation followed by \textit{dependent rounding}. 

There has been extensive work on developing dependent rounding techniques, that round the fractional solution in some correlated way to satisfy both the hard constraints and ensure some negative dependence amongst rounded variables that can result in concentration inequalities. For instance, the hard constraints might arise from an underlying combinatorial object such as a packing \cite{Brubach2017AlgorithmsTA}, spanning tree \cite{chekuri2010dependent}, or matching \cite{gandhiDependentRoundingIts2006} that needs to be produced. In our case, the rounded variables must satisfy both matching \eqref{eq:elli-caps}, \eqref{eq:ri-caps}, and barter constraints \eqref{eq:values-barter} (i.e., each agent gives the items of same total value as it received). \citet{gandhiDependentRoundingIts2006} developed a rounding scheme where the rounded variables satisfy the matching constraints along with other useful properties. Therefore, we adapt their rounding scheme (to satisfy matching constraints) followed by a careful rounding scheme that results in rounded variables  satisfying the barter constraints. 

Centralized barter exchanges are well-studied under various barter settings. For instance, \citet{abrahamClearingAlgorithmsBarter2007} showed that the bounded length \textit{edge-weighted directed cycle packing} is NP-Hard which led to several heuristic based methods to solve these hard problems, e.g., by using techniques of operations research \cite{constantino2013new,glorie2014kidney,plaut2016fast,carvalho2021robust}, AI/ML modeling \cite{mcelfreshImprovingPolicyConstrainedKidney2020,noothigattu2020axioms}. Recently several works focused on the fairness in barter exchange problems \cite{abbassi2013fair,fang2015randomized,klimentova2021fairness,farnadiIndividualFairnessKidney2021}. Our work adds to the growing body of research in theory and heuristics surrounding ubiquitous barter exchange markets.

\section{Outline of our contributions and the paper}

Firstly, we introduce the \BSV problem, a natural generalization of \textit{edge-weighted directed cycle packing} and show that it is NP-Hard to solve the problem exactly. Our main contribution is a randomized dependent rounding algorithm \BDR with provable guarantees on the quality of the \textit{allocation}. The following definitions help present our results. 
Suppose we are given an integral allocation $\{X_e\} \in \{0,1\}^{|E|}$, we define the \textit{net value loss} of each agent $i$ 
(i.e., the violation in the barter constraint \eqref{eq:values-barter}):
\begin{equation} \label{eq:Di-net-value}
    D_i := \sum_{b \in L_i} v_b X(b) - \sum_{a \in R_i} v_a X(a).
\end{equation}

Our main contribution is a rounding algorithm \BDR that satisfies both \textit{matching} (\eqref{eq:elli-caps} and \eqref{eq:ri-caps}) and \textit{barter} constraints (\eqref{eq:values-barter}) as desired in \textit{multi-agent} exchanges. Recall that the rounding algorithm \GKPSDR (indeed a pre-processing step of \BDR) rounds the fractional matching to an integral solution enjoying the properties mentioned in \cref{sec:prelim-gkps}. The main challenge in our problem is satisfying the \textit{barter constraint}. Here, a direct application of \GKPSDR alone can result in a worst case violation of $\sum_{a \in L_i} v_a$ on $D_i$, corresponding to the agent losing all their items and gaining none (see the example in the Appendix). However, our algorithm
\BDR rounds much more carefully to ensure, for each agent $i$, $D_i$ is at most $v_i^* := \max_{a \in L_i \cup R_i} v_a$, i.e., the most valuable item in $H_i \cup W_i$.
The two following theorems provide lower and upper bounds on tractable $D_i$ (i.e., \eqref{eq:Di-net-value}) guarantees for \BSV. 
\BDR on a bipartite graph $(L, R, E)$ is worst-case time $\Oh((|L| + |R|)(|L| + |R| + |E|))$
where $L, R = \Oh(|\mathcal{I}| n)$.
We view \cref{thm:bdr} as our main result.
\begin{theorem} \label{thm:bdr}
Given a \BSV instance,
\BDR is an efficient randomized algorithm achieving an allocation with optimal utility in expectation and where, for all agents $i$, $D_i < v_i^*$ with probability $1$ and $\E[D_i]=0$.
\end{theorem}
 \begin{theorem} \label{thm:bsv-hardness}
    Deciding whether a \BSV instance has a non-empty valid allocation with 
    $D_i=0$ for all agents $i$ is NP-hard, even if all item values are integers. 
\end{theorem}
Owing to its similarities to \GKPSDR, \BDR enjoys similar useful properties:
\begin{theorem} \label{thm:bdr-is-a-DR-alg}
    \BDR rounds $\{x_e\} \in [0,1]^{|E|}$ in the feasible region of \BSVLP into $\{X_e\} \in \{0,1\}^{|E|}$
    while satisfying \ref{prop:marginal}, \ref{prop:degree}, and \ref{prop:neg-corr}.
\end{theorem}

\paragraph{Outline of the paper}
In \cref{sec:bdr-alg} we describe \BDR (\cref{alg:modified-DR}), our randomized algorithm for
\BSV, and its subroutines \FCCC and \CCWalk in detail.
Next, we give proofs and proof sketches for \cref{thm:bdr,thm:bdr-is-a-DR-alg}.

\section{\texorpdfstring{\BDR}{BarterDR}: dependent rounding algorithm for \texorpdfstring{\BSV}{BarterSV}}
\label{sec:bdr-alg}
\paragraph*{Notation}
\BDR is a rounding algorithm that proceeds in multiple iterations, therefore we use a superscript $r$ to denote the value of
a variable at the beginning of iteration $r$.
An edge $e \in E$ is said to be \emph{floating} if $x_e^r \in (0,1)$. 
Analogously, let $E^r := \{ e \in E : x_e^r \in (0,1)\}$, $a \in L \cup
R$ is said to be a \emph{floating vertex} if $x^r(a) := \sum_{e \in N(a)} x_e^r \not \in \Z$ and
the sets of floating vertices are $L^r := \{ a \in L : x^r(a) \not \in \Z\}$ and
$R^r := \{ a \in R : x^r(a) \not \in \Z\}$.
$L(E^r) = \{ a \in L \mid \exists (a, b) \in E^r \wedge a \in L \}$. $R(E^r)$ is defined similarly. $G^r := (L(E^r), R(E^r), E^r)$. $K^r(a)$ denotes the connected component in $G^r$ containing vertex $a$.
Define $\kappa(i) := \{ a :  a \in L_i
\cup R_i \} $, for each $i \in [n]$, to be the set of participating vertices in
each barter constraint. We say two vertices $a, b \in L \cup R$ are \emph{partners} if
there exists $i \in [n]$ such that $a, b \in \kappa(i)$ and $a \ne b$. Note if $a$
and $b$ are partners, then they are distinct vertices corresponding to items
(owned or desired) of some common agent $i$. In iteration $r$, a vertex $a \in
\kappa(i)$ is said to be \emph{partnerless} if
$\kappa(i) \cap (L^r \cup R^r) = \{ a \}$;
i.e., $a$ is the only floating vertex in $\kappa(i)$.
We use the shorthand $a \sim b$ to denote $a$ and $b$ are partners.  
Edges and vertices not floating are said to be \emph{settled}:
once an edge $e$ (vertex $a$) is settled \BDR will not
modify $x_e$ ($x(a)$).
For vertices $a$ and $b$, $a \leadsto b$ denotes a simple path from $a$ to $b$.
Define $D_i^r$ to be $D_i$, as defined in \eqref{eq:Di-net-value}, but with variables
$\{x_e^r\}$ instead of $\{X_e\}$.
The fractional degree of $a \in L \cup R$ refers to $x^r(a)$.

Once an edge is settled, its value does not change. In the each iteration \BDR 
looks exclusively at the floating edges $E^r$ and the graph induced by them. Namely, $G^r := (L(E^r), R(E^r), E^r)$ where
$L(E^r) := \{ a \in L : \exists e \in E^r,\; e \in N(a)\}$ and $R(E^r)$ is defined analogously.
In each iteration, at least one edge or vertex becomes settled, i.e., 
$|E^r| + |L^r| + |R^r| > |E^{r+1}| + |L^{r+1}| + |R^{r+1}|$. Therefore \BDR
terminates in iteration $T$ where $|E^T|=0$ and $T \le |L| + |R| + |E|$.

\paragraph{Algorithm and analysis outline.}
\BDR
begins by making $G$ acyclic via the pre-processing step in \cref{sec:preprocessing}.
Next, \BDR proceeds as follows.
While there are floating edges find an appropriate sequence of paths $\Pc$ 
constituting a CCC or CCW (defined in \cref{sec:existence-cccs}).
The strategy for judiciously rounding $\Pc$ is fleshed out in \cref{sec:rounding-CCC}.
Finally, \cref{sec:analysis} concludes with proofs for \cref{thm:bdr,thm:bdr-is-a-DR-alg}.

\subsection{Pre-processing: remove cycles in \texorpdfstring{$G$}{G}}
\label{sec:preprocessing}
The pre-processing step consists of finding a cycle $C$ via depth-first search in the graph of floating edges and rounding $C$ via \GKPSDR until there are no more cycles.
Let $\{x_e^0\}_{e \in E}$ denote the LP solution and $\{x^1_e\}_{e \in E}$ denote the output of the pre-processing step. 
\BDR begins on $\{x^1_e\}_{e \in E}$.

\GKPSDR on cycles never changes fractional degrees, i.e., $\forall a \in L \cup R,\; x^0(a) = x^1(a)$. \cref{lemma:atleasttwo-unsat} is used to construct CCC's and CCW's, and it is the raison d'\^{e}tre for the pre-processing step. 
\begin{lemma}
The pre-processing step is efficient and gives $D_i^1 = 0$ for all agents $i$
with probability $1$.
\end{lemma}
\begin{proof}
    The pre-processing step runs \GKPSDR on $G$ until no cycles remain. \GKPSDR
    on a cycle guarantees that at the end of each iteration the fractional
    vertex degrees remain unchanged. Moreover, the full \GKPS algorithm takes
    time $O(|E| \cdot (||E| + |V|))$, bounding the pre-processing step's runtime
    which can be seen as \GKPS with an early stoppage condition.
\end{proof}
\begin{lemma} \label{lemma:unsat-vertices}
    Let $\{x_e\} \in (0,1)^{|E|}$ be a vector of floating edges over a bipartite
    graph $(U, V, E)$. Then the number of floating vertices in $(U, V, E)$ is
    not 1.
\end{lemma}
\begin{proof}%
    Let $u$ be the sole floating vertex, i.e., $x(u) \not\in \Z$, and, without
    loss of generality, let $u \in U$. For $S \subseteq U \cup V$, let $d(S) :=
    \sum_{s \in S} x_s$. Then $ d(U - \{u\}) + x_u = d(U) = \sum_{e\in E}x_e =
    d(V)$. But $u$ being the only floating vertex implies $d(U - \{u\}) + x(u)
    \not\in \Z$ and $d(V) \in \Z$.
\end{proof}
\begin{lemma} \label{lemma:atleasttwo-unsat}
Each connected component of $G^1$ has at least $2$ floating vertices.
\end{lemma}
\begin{proof}%
Fix an arbitrary connected component $K$ of $G^1$.
If $K$ has 0 floating vertices, then each vertex has
degree at least two because $\forall e \in E^r, x_e^r \in (0,1)$. This implies 
the existence of a cycle in $K$, but this cannot happen as the pre-processing step eliminates
all cycles. Applying \cref{lemma:unsat-vertices} to $K$, the only remaining
possibility is that $G^1$ has at least two floating vertices.
\end{proof}
\subsection{Construction of CCC's and CCW's via \texorpdfstring{\FCCC}{FindCCC}}
\label{sec:existence-cccs}
This section introduces CCC's and CCW's.
The definition of these structures facilitates rounding edges while
respecting the barter constraints each iteration.
The subroutines for constructing CCC's and CCW's, \FCCC and \CCWalk,
are described in \cref{alg:find-CCC,alg:CC-walk}. The correctness
of these subroutines, and thus the existence of CCC's and CCW's,
follows from \cref{lemma:CCC-construction}.
\begin{definition}
A \textit{connected component cycle} (CCC) is a sequence of $q \ge 1$ paths 
$\Pc = \langle s_1 \leadsto t_1, \dots, s_q \leadsto t_q \rangle$ such that, letting $V(\Pc) = \bigcup_{i \in [q]} \{ s_i, t_i\}$ be the paths' endpoint vertices,
\begin{enumerate}
    \item $\forall i \in [q]$, $t_{i} \sim s_{i+1}$ (taking $s_{q+1} \equiv s_1$), 
    \item $\forall a \in V(\Pc)$, $|V(\Pc) \cap \kappa(a)|=2$,
    \item $\forall i\in[q]$, $ s_i \leadsto t_i$ belongs to a unique connected component, and
    \item $\forall i \in [q]$, $s_i$ and $t_i$ are floating vertices.
\end{enumerate}
Instead, we have a \textit{connected component walk} (CCW) if criteria 3) and 4) are met but 1) and 2) are relaxed to:
1) $\forall i \in [q-1]$, $t_{i} \sim s_{i+1}$ and $s_1$ and $t_q$ are partnerless;
and 2) $\forall a \in V(\Pc) - \{ s_1, t_q\}, |V(\Pc) \cap \kappa(a)|=2$. 
\end{definition}
\begin{figure}
    \centering
    \includegraphics[width=.5\textwidth]{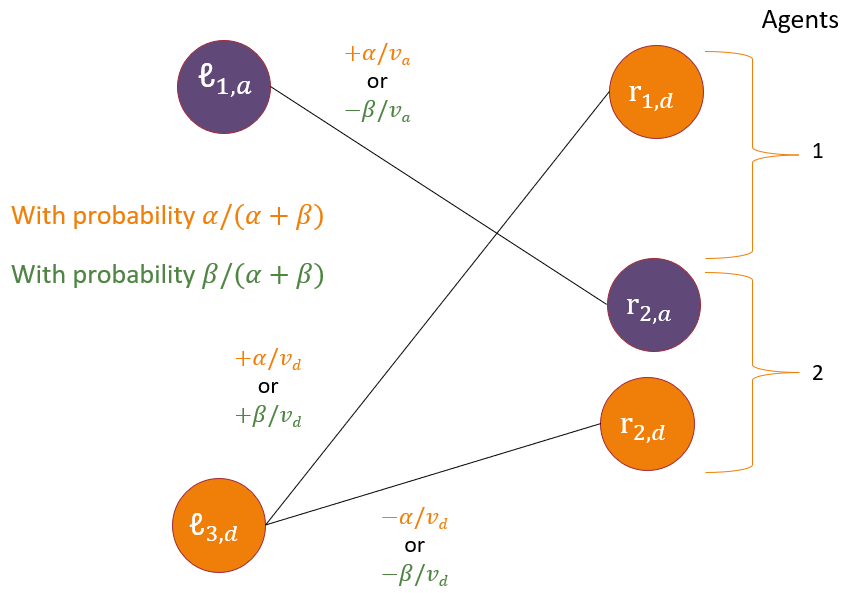}
    \caption{An example CCC $\Pc = \left< \ell_{1,a} \leadsto r_{2,a}, r_{2,d} \leadsto r_{1,d}\right>$ based on the VBM from \cref{fig:vbm}. All edges and path endpoint vertices, i.e., $\ell_{1,a}$, $r_{2,a}$, $r_{1, d}$, and $r_{2,d}$, must be floating. The vertex $\ell_{3,d}$. The rounding step proceeds in one of two ways denoted by {\color{orange}orange} and {\textcolor{teal}{green}} text, chosen at random. With the probability $\alpha / (\alpha + \beta)$, displayed in orange, the orange modifications to $x_e$'s take place. Otherwise, with probability $\beta / (\alpha + \beta)$, the green modifications take place. In the orange rounding event agent 1 gives away an additional $v_a \alpha / v_a = \alpha$ value and gains an additional $v_d \alpha / v_d = \alpha$ value therefore $D_i$ for agent 1 does not change after this rounding step. It is easy to see the same occurs for both agents, in either -- orange or green -- rounding event.}
    \label{fig:ccc}
\end{figure}
Recall that a rounding iteration $r$ is fixed so whether a vertex is floating or partnerless is well-defined.
When $\Pc$ is rounded the set of vertices whose fractional degrees change is precisely $V(\Pc)$.
Requirements 1 and 2 of a CCC say $t_i$ and $s_{i+1}$ are partners \emph{and}
they do not have any other partner vertices in $V(\Pc)$. 
Comparably, for CCW's these requirements 
imply the same for all vertices but the ``first'' and ``last,'' which are partnerless. 
Therefore, for CCC's and CCW's the vertices in $V(\Pc)$ respectively appear in $q$ and $q+1$ distinct barter constraints.
The requirements in the definitions of CCC and CCW come in handy during the analysis because:
each path belongs to a different connected component hence they are vertex and edge disjoint; if a barter constraint has exactly two vertices in $V(\Pc)$ then these vertices' fractional degree changes can be made to cancel each other out in the barter constraint; and floating vertices ensure paths can be rounded in a manner analogous to \GKPSDR. 
For comparison, \GKPSDR also needed paths with floating endpoints, but maximal paths always have such endpoints whereas the paths of $\Pc$ need not be maximal. Consequently, the requirement that paths of $\Pc$ have floating endpoints must be imposed.
\begin{algorithm}[ht]
    \DontPrintSemicolon
    \caption{\BDR}
    \label{alg:modified-DR}
    \KwIn{$\{x_e^1\} \in [0,1]^{|E|}$, corresponding to $G^1=(L(E^1),R(E^1),E^1)$; i.e., the output of the pre-processing described in \cref{sec:preprocessing}}
    $r \gets 1$\;
    \While{$E^r \ne \emptyset$} {
        $\Pc \gets$ CCC or CCW returned by \FCCC in $G^r$\;
        Round $\Pc$ as described in \cref{sec:rounding-CCC} yielding $G^{r+1}$ and $\{ x_e^{r+1}\}$\;
        $r \gets r+1$\;
    }
    \Return{$\{x_e^r\} \in \{0, 1\}^{|E|}$}
\end{algorithm}
\begin{algorithm}[ht]
    \DontPrintSemicolon
    \caption{\textsc{FindCCC}}
    \label{alg:find-CCC}
    \KwIn{$G^r=(L(E^r),R(E^r),E^r)$}
    \KwOut{CCC or CCW \CCC}
    $s_1,t_1 \gets$ distinct floating vertices in some connected component $C$ of $G_r$\;
    $(O_1, \sigma_1) \gets$ \CCWalkp{t_1} \;
    \uIf{$\sigma_1 = $ "CCC"}{
        \Return{$O_1$}
    }
    $(O_{2}, \sigma_{2}) \gets$ \CCWalkp{s_1} \;
    \uIf{$\sigma_2 = $ "CCC"}{
        \Return{$O_2$}
    }
    Let $O_1 = (t_1, s_2, t_2, \dots, s_q, t_q)$ and
    $O_2 = (t'_1, s'_2, t'_2, \dots, s'_{q'}, t'_{q'})$\;
    If $O_1$ and $O_2$ ``cross,'' resolve $O_1$ and $O_2$ into a CCC and return it; see \cref{sec:existence-cccs} \label{algline:uncross-ccw}\;
    \Return{$\left< t'_{q'} \leadsto s'_{q'}, t'_{q'-1}, s'_{q'-1}, \dots,
    s'_2 \leadsto t'_2, t'_1 \leadsto t_1, s_2 \leadsto t_2, \dots, s_q \leadsto t_q \right> $} \tcp*{CCW}
\end{algorithm}
\begin{algorithm}[ht]
    \DontPrintSemicolon
    \caption{\CCWalkp{a}}
    \label{alg:CC-walk}
    \KwIn{$G^r=(L(E^r),R(E^r),E^r)$, the walk's starting vertex $a$}
    \KwOut{A CCC or half of a CCW and a string indicating whether a CCC was returned}
    $V \gets (t_1)$, letting $t_1 := a$  \tcp*{ordered list of path endpoints}
    $S \gets \{C_1\}$ where $C_1 = K^r(a)$ \tcp*{$S$ is the set of seen CC's}
    $i \gets 2$\;
    \While{True} {
        \uIf{$t_{i-1}$ is partnerless} {
        \Return{$(V, \text{"CCW"})$} \tcp*{Return half of a CCW}
        }
        $s_i \gets$ partner of $t_{i-1}$\;
        $C_i \gets$ connected component containing $s_i$\;
        \uIf(\tcp*[f]{$C_i$ was previously visited}){
        \label{algline:revisit-cc}
        $C_i \in S$} {
        $C_i = C_k$ for some $k < i$ so let $s_k' := s_i$\;
        \Return{
        $(\left< s'_k \leadsto t_k ,\dots, s_{i-1} \leadsto t_{i-1} \right>, \text{"CCC"})$ }
        }
        \label{algline:another-floating-vertex}
        $t_i \gets$ floating vertex in $C_i$ distinct from $s_i$\;
        $V \gets V \oplus (s_i)$, where $\oplus$ denotes sequence concatenation\;
        \uIf(\tcp*[f]{$t_i$ already has a partner in $V$})
        { \label{algline:revisit-partner} 
        $\exists b \in V,\; b \sim t_i$
        }
        {
        It must be that $b$ already had a partner $c \in V$. WLOG, $b=t_{k-1}$ and $c=s_{k}$, some $k \le i$\;
        \Return{$(\left< s_k \leadsto t_k, \dots, s_i \leadsto t_i \right>, \text{"CCC"})$}
        }
        $V \gets V \oplus (t_i)$\;
        $i \gets i+ 1$,\quad $S \gets S \cup \{ C_i \}$
    }
\end{algorithm}
\paragraph{Uncrossing the half-CCWs.} \label{par:uncross-half}
We define what we mean by "crossing" half-CCW's in \FCCC~\cref{algline:uncross-ccw}
and how to resolve this into a CCC.
Using $O_1$ and $O_2$ build
$\bar{V} := \left< s_q', t_{q-1}', \dots,s_2', t_1', t_1, s_2, \dots, s_q \right>$. $\bar V$ can be seen as the sequence of path endpoints (i.e., $V$ in \CCWalk) resulting from a run of \CCWalkp{s_q'}, which possibly did not stop when it should have returned a CCC. By the half-CCW's "crossing" we mean that in some iteration of the while-loop of \CCWalk either a connected component is revisited or $t_i$ was partners with a vertex previously visited. But these cases are precisely \cref{algline:revisit-cc,algline:revisit-partner} from \CCWalk where a CCC is resolved and returned. 
\begin{lemma}\label{lemma:CCC-construction}
    If $G^r$ has no cycles, \FCCC efficiently  returns a CCC or CCW.
\end{lemma}
\begin{proof}[Proof of \cref{lemma:CCC-construction}]
$G^r$ for $r\ge1$ is guaranteed to be acyclic by the pre-processing step. Recall, this acyclic property is used by \cref{lemma:atleasttwo-unsat} to guarantee each CC has at least two floating vertices. This ensures \cref{algline:another-floating-vertex} of \FCCC is well defined.

Let $V_r$ be the \emph{sequence} of vertices $V$ at the beginning of iteration $r$, and let $V_r - a$ denote $V_r$ without some vertex $a$.
Like in \CCWalk, ``$\oplus$'' denotes sequence concatenation.
Let $a_r$ and $z_r$ be the first and last vertices of $V_r$; note $a_r$ does not change over iterations.
We prove the correctness of \CCWalk with the aid of the following loop invariants
maintained at the beginning of each iteration $r$ of the while-loop.
\begin{enumerate}[label=\textbf{(I\arabic*)}]
    \item $z_r$ has no partners in $V_r$.
    \label{invar:ccwalk-last-no-partner}
    \item $\forall b \in V_r - z_r$, $b$ has exactly one partner in $V_r$.
    \label{invar:ccwalk-not-last-partners}
    \item  
    $a_r$ is the only vertex from $V_r$ in $K(a_r)$.
    \label{invar:ccwalk-first-cc}
    \item $\forall b \in V_r - a_r$, there are exactly two vertices 
    from $V_r$ contained in $K(b)$.
    \label{invar:ccwalk-two-per-cc}
\end{enumerate}
Proceed by induction. When $r=1$, $V_r = \seq{a}$ so $a_r = a = z_r$ so all invariants
are (vacuously) true. Let $P(k)$ be the predicate saying all invariants hold at the beginning of iteration $k \ge 1$. We assume $P(k)$ and show $P(k+1)$.

If there is an iteration $k+1$, then \CCWalk did not return during iteration $k$ and must have added $\seq{s_k, t_k}$ to $V_k = \seq{t_1, s_2, t_2, \dots, s_{k-1}, t_{k-1}}$.
If $z_{k+1} = t_k$ had a partner in $V_k \oplus \seq{s_k}$ then iteration $k$ would have been the last as a CCC would have been returned. Therefore \ref{invar:ccwalk-last-no-partner} holds at the beginning of iteration $k+1$.

By $P(k)$, $\forall b \in V_k - t_{k-1}$, $b$ has exactly one partner in $V_k$ and $t_{k-1}$ has zero partners in $V_k$. By construction $s_k$ is selected to be the partner of $t_{k-1}$ so now $t_{k-1}$ and $s_k$ have exactly one partner each in $V_{k+1}$. Therefore \ref{invar:ccwalk-not-last-partners} holds at the beginning of iteration $k+1$.

If $V_{k+1}$ were to not meet \ref{invar:ccwalk-first-cc} then it must mean that
either $s_k \in K(a_k)$ or $t_k \in K(a_k)$. But in this case the connected component $K(a_k)$ was revisited during iteration $k$ and a CCC would have been returned. Therefore, \ref{invar:ccwalk-first-cc} holds.

By construction $K(s_k) = K(t_k)$. Moreover, $\forall b \in V_k,\; K(s_k) \ne K(b)$
otherwise $K(b)$ was revisited and iteration $k$ would have been the last. 
Therefore, \ref{invar:ccwalk-two-per-cc} continues to hold.

Moreover note the while-loop runs at most $\Oh(|L| + |R|)$ many times since revisiting
a connected component of $G^r$ causes the function to return.

Next we leverage the loop invariants to prove \CCWalk returns valid CCC's.
First observe that by construction of $V$, Properties 1) and 4) of a CCC are always immediate.
\CCWalk returns CCC's when a connected component is revisited or when the added $t_r$
already has partners present in $V$. We inspect both return cases to ensure indeed a valid CCC is returned.
Fix some iteration $r \ge 1$. 

Suppose a connected component $C_k$, $k < r$ is revisited.
Then $\left< s'_k \leadsto t_k, \dots, s_{r-1} \leadsto t_{r-1} \right>$ is returned.
By \ref{invar:ccwalk-last-no-partner}, $t_{r-1}$ and $s'_k$ are each others only partners.
Recall by construction $t_k \sim s_{k+1}$, $t_{k+1} \sim s_{k+1}, \dots,$ and $ s_{r-1} \sim t_{r-1}$,
so by \ref{invar:ccwalk-not-last-partners} it follows that each of $t_k, s_{k+1}, t_{k+1}, \dots, s_{r-1}, t_{r-1}$ has exactly one partner amongst themselves. 
Therefore, Property 2) of CCC's holds. It remains to check Property 3).
By \ref{invar:ccwalk-two-per-cc} each of $s_p \leadsto t_p$, for $k \le p < r$ belongs to a unique connected 
component. Therefore, $s'_k \leadsto t_k$, too, must belong to a unique connected component different from each $s_p \leadsto t_p$; otherwise there was a connected component containing $s_p$, $t_p$, and $s_k$ contradicting one of \ref{invar:ccwalk-first-cc} and \ref{invar:ccwalk-two-per-cc}
(depending on whether $k=1$ or $k>1$).

Instead suppose a CCC is returned because $t_r$ had a partner $b \in V_r \oplus \left< s_r\right>$. Let $b$ be the last vertex in $V_r \oplus \left<s_r\right>$ such that $b \sim t_r$.
If $b = s_r$ then $\left< s_r\leadsto t_r\right>$ is clearly a CCC. So suppose $b \in V_r$. It must be that $b = s_p $ for some $p < r$; otherwise $b$ is not the last such vertex. So focus on proving $\left< s_p \leadsto t_p, \dots, s_r \leadsto t_r \right>$ is a CCC. Property 2) follows because
$t_{r-1}$ and $s_r$ are each other's only partners by \ref{invar:ccwalk-last-no-partner}; $t_r$ and $s_p$ are each other's only partners because by \ref{invar:ccwalk-not-last-partners} $s_p$ previously had only partner $t_{p-1}$ but we have cut it out
from the CCC; and the rest of the pairs have unique partners by \ref{invar:ccwalk-not-last-partners}. Lastly, Property 3) holds because $K(s_r)$ was not a revisited CC
so $s_r$ and $t_r$ share a unique CC, and the rest of the path endpoints belong to distinct CC's by \ref{invar:ccwalk-two-per-cc}.

The last remaining case is that where the two half-CCW's overlap. This can be resolved into a CCC in the manner already described in the main text under the paragraph \emph{Uncrossing the half-CCW's} of 
\cref{sec:existence-cccs}.

\paragraph{Runtime of \FCCC}
We conclude with comments about the runtime of \FCCC.
We can build a hash-map mapping vertices to a set of all their floating partners. 
This hash-map can be constructed in time $\Oh(|L| + |R|)$.
Similarly, we can build a set to keep track of visited connected components.
Finding the partner $s_i$ of $t_{i-1}$ can be done in $\Oh(1)$ time by checking the hash-map, and finding the vertex $t_i$ in $K(s_i)$ can be done by starting
a depth first search from $t_i$ until a floating vertex is reached.
After rounding the CCC/CCW, remove vertices that became settled from their respective sets of floating partners.
Altogether, the depth first searches starting from each $t_i$ altogether visit each vertex and edge at most $\Oh(1)$ times before returning and each vertex is removed from the hash-map's set of floating partners at most once. 
Therefore, \CCWalk runs in time $\Oh(|L|+|R|)$.
\FCCC calls \CCWalk $\Oh(1)$ times and resolving the two-half CCW's can be thought of as another run of \CCWalk, as argued above. Therefore, \FCCC finishes in time
$\Oh(|L| + |R|)$.
\end{proof}

\subsection{Rounding CCC's and CCW's} \label{sec:rounding-CCC}
\subsubsection{Roundable colorings}
Now we shed light on what we mean by carefully rounding the paths of the CCC/CCW $\Pc$.
But first we build some intuition.
Focus on $t_p$ and $s_{p+1}$ for some fixed $1 \le p \le q$ in case of a CCC (or $p < q$ in the case of a CCW). Since $t_p \sim s_{p+1}$, whatever
rounding procedure we use, we want the relative signs of the changes to 
$x^r(t_p)$ and $x^r(s_{p+1})$ to depend on whether $t_p$ and $s_{p+1}$ fall
on the same or different sides of $G$ (these sides being ``left'' and ``right'' corresponding to vertex sets $L$ and $R$; equivalently, left and right of ``$=$'' in \eqref{eq:values-barter}). This way \eqref{eq:values-barter} is preserved after rounding. Likewise, the
magnitudes of fractional degree changes to $t_p$ and $s_{p+1}$ 
must be balanced depending on $v_p$ and $v_{p+1}$ so that \eqref{eq:values-barter} is preserved for $i$ corresponding to $t_p, s_{p+1} \in \kappa(i)$. 
Intuitively, we round $\Pc$ carefully \eqref{eq:values-barter} by ensuring i) the signs of the changes depend on which side of the equation tehy appear on, and ii), the magnitudes of said changes are weighted by the values of the items they represent. We make this precise via \emph{roundable colorings}.
If $a$ and $b$ are vertices belonging to different sides of the graph, we say $a \perp b$;
otherwise we say $a \not\perp b$.
\begin{definition}[Roundable coloring]
The CCC $\Pc=$\CCC has a \textit{roundable coloring} if there exists $\cf: V(\Pc) \to \{-1,1\}$ such that i) for all $i \in [q]$, $\cf(s_i) = \cf(t_i)$
if and only if $s_i \perp t_i$; and ii) for all $i \in [q]$, $\cf(t_i) = \cf(s_{i+1})$ if and only if $t_i \perp s_{i+1}$.
A roundable coloring for a CCW is defined the same way except ii) 
becomes $\forall i \in [q-1]$, $\cf(t_i) = \cf(s_{i+1})$ if and only if $t_i \perp s_{i+1}$.
\end{definition}
Property i) will ensure $a,b \in V(\Pc)$ see same-sign fractional degree change if and only if $\cf(a) = \cf(b)$, which happens if and only if they appear on the same side \eqref{eq:values-barter}.
Property ii) is
equivalent to \cref{remark:gkps-path-length} and verifies
each $s_i \leadsto t_i$ is roundable in \GKPSDR-inspired manner.
\begin{lemma} \label{lemma:roundable-coloring}
    Every CCC and CCW admits an efficiently computable roundable coloring $\cf$.
\end{lemma}
\begin{proof}[Proof of \cref{lemma:roundable-coloring}]
    Consider the following heuristic to find a roundable coloring of $\Pc = \mCCC$. Assign an arbitrary color to $s_1$, say $\cf(s_1)=1$.
    Because $\cf$ only has two colors, this immediately determines the color of $t_1$,
    which depends on whether $s_1$ and $t_1$ are same-side vertices.
    Again, this immediately determines the color of $s_2$,
    which in turn determines the color of $t_2$, and so on. 
    
    If $\Pc$ is a CCW then $s_1$ and $t_q$ are partnerless so
    their colors do not affect the first property. The vertices are colored
    in the order $s_1, t_1, s_2, t_2, \dots, s_q, t_q$ ensuring that both properties hold.
    Therefore, this scheme produces a roundable coloring  for CCW's in
    time $\Oh(|L|+|R|)$.
    
    Instead suppose $\Pc$ is a CCC. It only remains to check that $\cf(t_q) = \cf(s_1)$.
    We now verify this. Observe that the greedy algorithm ensures
    \[ \forall i \in [q],\; \cf(t_i) = \cf(s_i) \cdot (-1)^{\mathbbm{1}(t_i \perp s_i)} 
    \]
    \text{and}
    \[
    \forall i \in [q-1],\; \cf(s_{i+1}) = \cf(t_i) \cdot (-1)^{\mathbbm{1}(t_i \not\perp s_{i+1})} \]
    where $\mathbbm{1}(\cdot)$ equals 1 if ``$\cdot$'' is true and 0 otherwise; this is a slight
    abuse of notation since $\perp$ is a relation but we are treating it as a boolean function.
    Expanding by repeated application of the above observations, we have
    \begin{align*}
        \cf(t_q) &= \cf(s_1) \cdot (-1)^{\sum_{i \in [q]}\ind(t_1 \not\perp s_i) + 
        \sum_{i \in [q-1]} \ind(t_i \not\perp s_{i+1})} \\
        &= \cf(s_1) \cdot (-1)^{p + \psi_{q-1}}
    \end{align*}
    where $p$ is the number of same-side paths and $\psi_{q-1}$ is the number of same-side
    partners not counting the pair $t_q$ and $s_1$. Then ensuring we have a roundable coloring reduces to
    ensuring that $t_q \perp s_1 \implies p + \psi_{q-1}$ is even and $t_q \not\perp s_1 \implies
    p + \psi_{q-1} $ is odd. Letting $\psi$ be the total number of same-side partners, the above is equivalent to
    asking $p+\psi$ be even, which we now prove.
    
    Let $d$ be the number of different-side paths, and let $c_L$, $c_R$, $c_{LR}$ respectively be the
    number of left-left, right-right, and left-right partner pairs. 
    So,
    \begin{equation} \label{eq:num-paths-and-partners}
    p + d = q = \psi + c_{LR}.
    \end{equation}
    Let $n_L $ be the number of left vertices in the CCC. Clearly, $n_L = 2c_L + c_{LR}$.
    Look at $n_L - d$, this is the number of left vertices remaining after removing the 
    different-side paths in \CCC. Since these vertices must be covered by same side paths
    we must have $n_L - d$ even. Then, with
    all congruences taken modulo $2$,
    $$ 0 \equiv n_L - d = 2c_L + c_{LR} - d \equiv c_{LR} - d.$$
    Plugging the above into \eqref{eq:num-paths-and-partners} gives
    $p = \psi + c_{LR} - d \equiv \psi$.
    Therefore, $\psi + p \equiv \psi + \psi \equiv 0$.
\end{proof}

\subsubsection{Using roundable colorings}
 Roundable colorings will ensure the sings of changes are as desired. We now introduce notation to use the roundable colorings and ensure that the magnitudes
 of said changes is done in proportion to the value of items involved.  
 
Although the notation used next is cumbersome, the intuition is to fix $\alpha, \beta > 0$ 
``small enough'' that all edge variables stay in $[0,1]$ and vertex fractional degrees stay within their current ceilings and floors but ``large enough'' 
that at least one edge or vertex is settled.
First, fix the roundable coloring $\cf$, which is possible per \cref{lemma:roundable-coloring}.
Next, decompose each path $s_i \leadsto t_i$  into alternating matchings $M_{-1}^i$
and $M_{1}^i$ such that $\forall a \in \{s_i,t_i\},\; \exists e \in M_{\cf(a)}^i $ such that $e \in N(a)$; property ii) of $\cf$ guarantees this is possible.
In other words, vertex $a\in \{s_i, t_i\}$ is present in $M_{\cf(a)}^i$.
For readability drop the $r$ superscripts briefly and let
\begin{align} \label{eq:Gamma-1}
    \Gamma_{-1}^i(\gamma) \equiv& \bigvee_{e \in M_{-1}^i} (x_e + \gamma = 1) \vee 
    \bigvee_{e \in M_1^i} (x_e - \gamma = 0) 
    \nonumber \\
    &\vee
    \bigvee_{a \in \{s_i, t_i\}} (\cf(a) = -1 \implies x(a) + \gamma = \lceil x(a) \rceil) 
    \phantom{\Gamma_{-1}^i(\gamma)\equiv}
    \nonumber \\
    &\vee  \bigvee_{a \in \{s_i, t_i\}} (\cf(a) = 1 \implies x(a) - \gamma = \lfloor x(a) \rfloor),
\end{align}
    \text{and, symmetrically,}
\begin{align} \label{eq:Gamma1}
    \Gamma_{1}^i(\gamma) \equiv& \bigvee_{e \in M_{1}^i} (x_e + \gamma = 1) \vee
    \bigvee_{e \in M_{-1}^i} (x_e - \gamma = 0)
    \nonumber \\
    &\vee  \bigvee_{a \in \{s_i, t_i\}} (\cf(a) = 1 \implies x(a) + \gamma = \lceil x(a) \rceil) \nonumber \\
    \phantom{\Gamma_{1}^i(\gamma) \equiv}&\vee \bigvee_{a \in \{s_i, t_i\}} (\cf(a) = -1 \implies x(a) - \gamma = \lfloor x(a) \rfloor).
\end{align}
Finally, the magnitudes fixed (in analogy to \cref{sec:prelim-gkps}) are
\begin{equation} \label{eq:bdr-alpha-beta}
\alpha := \min\left\{ \gamma > 0 : \bigvee_{i \in [q]} \Gamma_{-1}^i \left( \frac{1}{v_i} \gamma \right) \right\},
\; 
\beta := \min\left\{ \gamma > 0 : \bigvee_{i \in [q]} \Gamma_{1}^i \left( \frac{1}{v_i} \gamma \right) \right\}.
\end{equation}
Both $\alpha$ and $\beta$ are well defined as they are the minima of non-empty finite sets.
The update proceeds probabilistically as follows: 
$\forall i \in [q], \forall e \in s_i \leadsto t_i$,
\begin{equation} \label{eq:round-event-A}
\text{w.p.}\; \frac{\beta}{\alpha + \beta},\;
x_e^{r+1} = 
\begin{cases}
x_e^r + \frac{1}{v_i} \alpha, &e \in M_{-1}^i\\
x_e^r - \frac{1}{v_i} \alpha, &e \in M_{1}^i
\end{cases}
;
\end{equation}
\begin{equation}
\text{else, w.p.}\; \frac{\alpha}{\alpha + \beta},\;
\;
x_e^{r+1} = 
\begin{cases}
x_e^r - \frac{1}{v_i} \beta, &e \in M_{-1}^i\\
x_e^r + \frac{1}{v_i} \beta, &e \in M_{1}^i
\end{cases}.
\end{equation} 
At a high level $\alpha$ and $\beta$ are large enough that every rounding iteration at least one vertex or edge is settled. Adding the $\pm \frac{1}{v_i}$ terms ensures term appearing in \eqref{eq:values-barter} will cancel out nicely, as seen in the proof of \cref{lemma:Di-preserved}.

\subsection{Algorithm analysis} \label{sec:analysis}
\paragraph{Proof sketch of \cref{thm:bdr-is-a-DR-alg}}
Except for \ref{invar:degree}, the proofs for the invariants are almost identical to those in \cite{gandhiDependentRoundingIts2006}. This is because \BDR is crafted so as to be similar to \GKPSDR 
in the ways necessary for this analysis to carry over. Owing to proofs' similarities, we defer their full treatment to \cref{sec:proof-of-bdr-is-a-DR-alg}.
The proof proceeds via the following invariants maintained at each iteration $r$ of \BDR.
\begin{enumerate}[label=\textbf{(J\arabic*)}]
    \item $\forall e\in E$, $\E[x_e^r] = x_e^0$. \label{invar:marginal}
    \item $\forall a \in L \cup R$ and with probability 1, $\lfloor x^0(a) \rfloor\le x^r(a) \le  \lceil x^0(a) \rceil$. \label{invar:degree}
    \item $\forall a \in L \cup R,\; \forall S \subseteq N(a),\; \forall c \in \{0,1\}$,
    $\E[\prod_{e \in S}x_e^{r+1}] \le \E[\prod_{e \in S} x_e^r]$.
    \label{invar:neg-corr}
\end{enumerate}
The main place where our proof deviates results from \BDR choosing $\alpha$ and $\beta$ differently.
In particular, \GKPSDR selects a single path and at least one of its edges becomes settled each iteration. In \BDR the guarantee says, among all paths of the fixed CCC/CCW, at least one edge \emph{or} vertex becomes settled each iteration.

\begin{lemma} \label{lemma:bdr-opt-obj-and-EDi}
\BDR achieves optimal objective in expectation and $\forall i \in [n],\; \E[D_i]=0$.
\end{lemma}
\begin{proof}[Proof of \cref{lemma:bdr-opt-obj-and-EDi}]
    Given a \BSV instance, let $\OPTIP$ and $\OPTLP$ be the optimal objectives of the corresponding
    IP \eqref{eq:barter-IP} and the corresponding \BSVLP.
    Let $\{X_e^*\}_{e \in E}$ and $\{x_e^*\}_{e \in E}$ be optimal solutions to the IP and LP, respectively. Then 
    $\OPTIP = \sum_{e \in E} w_e X_e^* \le  \sum_{e \in E} w_e x_e^* = \OPTLP$.
    Per \cref{thm:bdr-is-a-DR-alg}, \BDR satisfies \ref{prop:marginal} when rounding $\{x_e^*\}_{e \in E}$ to $\{X_e\} \in \{0,1\}^{|E|}$. Therefore,
    $\E[\sum_{e \in E} w_e  X_e ] = \sum_{e \in E} w_e \E[ X_e ] =
    \sum_{e \in E} w_e x_e^* = \OPTLP$. 

    By the linearity of expectation
    and \ref{prop:marginal}, 
    \begin{align*}
    \E[D_i] &= \E[\sum_{a \in L_i} X(a) v_a - \sum_{b \in R_i} X(b)v_b ] \\
    &=\sum_{b \in L_i} \E[X(b)]v_b - \sum_{a \in R_i} \E[X(a)] v_a \\
    &= \sum_{b \in L_i} x^*(b)v_b - \sum_{a \in R_i} x^*(a) v_a = D_i^0 = 0.
    \end{align*}
    The last equation follows because $\{x^*_e\}_{e\in E}$ satisfies
    \eqref{eq:values-barter} as argued in the proof of \cref{thm:bdr-is-a-DR-alg}.
\end{proof}
\begin{lemma} \label{lemma:Di-preserved}
    If $D_i^r = 0$ and there exists distinct floating $a,b \in \cf(i)$, then $D_i^{r+1}=0$.
\end{lemma}
\begin{proof}[Proof of \cref{lemma:Di-preserved}]
    If no vertex from $L_i \cup R_i$ appears in the CCC/CCW's endpoints
    $V(\Pc) := \bigcup_{i\in [q]} \{s_i, t_i\}$
    then we are done.
    So suppose $a \in L_i \cup R_i$ and $a \in V(\Pc)$ in this $r$-th rounding iteration. 
    By assumption there exists another floating vertex, namely
    $b \in L_i \cup R_i$ in iteration $r$ when \CCC was constructed. 
    Therefore, $a$ is not partnerless hence it cannot be the endpoint of a CCW so
    there exists $b \in V(\Pc)$ such that $a \sim b$.
    Moreover, by property 2 of the definition of a CCC/CCW, said $b$ is unique.
    Therefore, $a$ and $b$ are the \emph{only} vertices in $V(\Pc)$ affecting $D_i$
    this iteration $r$;
    i.e., $\forall d \in (L_i \cup R_i) - \{a,b\},\; x^r(d) = x^{r+1}(d)$.
    Since $a \sim b$, 
    we may assume without loss of generality that $a=t_k$ and $b=s_{k+1}$ for some $k \in [q]$ (or $k \in [q-1]$ for a CCW); recall $s_{q+1} \equiv s_1$.
    We know $\cf(a) = \cf(b)$ if and only if $a$ and $b$
    belong to opposite graph sides where $\cf$ is the valid coloring function corresponding to $\Pc$,
    which can be fixed efficiently per \cref{lemma:roundable-coloring}.
    
    Consider the two possible
    rounding events, described in \eqref{eq:round-event-A}. Call these events $\theta_1$ and $\theta_2$.
    Suppose $a$ and $b$ are opposite-side vertices, hence $\cf(a) = \cf(b)$. Focus on event $\theta_1$ (the proof for $\theta_2$ is exactly the same but replacing $\alpha$ with $-\beta$). Under event $\theta_1$ we have
    \begin{equation*}
    x^{r+1}(a) = x^{r}(a) - \cf(a) \frac{1}{v_a} \alpha
    \quad \text{and} \quad
    x^{r+1}(b) = x^{r}(b) - \cf(b) \frac{1}{v_b} \alpha.
    \end{equation*}
    A moment's notice here shows that $x^{r+1}(a), x^{r+1}(b) \in [0,1]$ per the definitions of 
    \cref{eq:Gamma-1,eq:Gamma1,eq:bdr-alpha-beta}.
    Note the factor of ``$-\cf(a)$'' appears because $a$ belongs to $M^i_{\cf(a)}$ for some $i$.
    Conveniently, this leaves us with 
    \begin{align}
        x^{r+1}(a)v_a - \cf(a)\cf(b) x^{r+1}(b)v_b 
        &= x^r(a) v_a - \cf(a)\cf(b) x^{r}(b)v_b
        -\cf(a) \alpha +\cf(a) \alpha \\
        &= x^r(a) v_a - \cf(a)\cf(b) x^{r}(b) v_b, \label{eq:simpler-xa-xb-diff}
    \end{align}
    using the fact $\cf(b)\cdot \cf(b) =1$.
    Without loss of generality let $a \in L_i$. 
    Therefore, expanding $D_i^{r+1}$:
    \begin{equation} 
        D_i^{r+1} = \sum_{s \in L_i} x^{r+1}(s) v_t - \sum_{t \in R_i} x^{r+1}(t) v_t.
    \end{equation}
    Having fixed $a \in L_i$, we know $\cf(a)\cf(b) = 1$ if and only if
    $b \in R_i$.
    Thus, take out $x^{r+1}(a)$ and $x^{r+1}(b)$ from the sums and substitute \eqref{eq:simpler-xa-xb-diff} to have
    \begin{equation} \label{eq:change-a-and-b}
        \sum_{s \in L_i - \{a, b\}} x^{r+1}(s) v_t  - \sum_{t \in R_i - \{b\}} x^{r+1}(t) v_t 
        + x^r(a) v_a - \cf(a)\cf(b) x^{r}(b) v_b.
    \end{equation}
    Now observe that $x^{r+1}(p)=x^r(p)$ for all $p \in (L_i \cup R_i ) - \{a,b\}$.
    Moreover, $b \in R_i$  if and only if $\cf(a)\cf(b)=1$, so
    we can reabsorb the terms ``$x^r(a) v_a$'' and ``$x^r(b) v_b$'' into their
    respective summations; thus yielding
    $ \sum_{s \in L_i} x^{r}(s) v_t  - \sum_{t \in R_i} x^{r}(t) v_t$.
    But this is precisely $D_i^r$, which we've assumed to be 0.
\end{proof}

\begin{proof}[Proof of \cref{thm:bdr}]
It is straightforward to check that after solving \BSVLPcap there are at most $|E|$ floating edges.
Each iteration of the pre-processing step finds a cycle, say using depth-first-search, and rounds said cycle in time $\Oh(|L| + |R|)$ with at least one edge being settled every time a cycle is rounded. 
Therefore, the pre-processing step takes time at most $\Oh(|E| \cdot (|L| + |R|))$.
Similarly, \FCCC takes time $\Oh(|L| + |R|)$ to find and round a CCC or CCW.
Each iteration a CCC/CCW is rounded either one edge or vertex becomes settled. Therefore
\BDR runs in time $\Oh((|L| + |R|) \cdot (|L| + |R| + |E|))$.

Let $D_i^r$ be $D_i$ like in \eqref{eq:Di-net-value} but with variables $x_e^r$
instead of $X_e$.
Then \cref{lemma:Di-preserved} guarantees that for each agent $i$, 
$D_i^r=0$ implies $D_i^{r+1}=0$ until $L_i \cup R_i$ has exactly one 
floating vertex (if this happens at all).
This means if in some iteration the number of floating vertices in $L_i \cup R_i$ went from at least 2 to 0, 
then $D_i=0$ by the degree preservation invariant
\ref{invar:degree}, proved in the proof of \cref{thm:bdr-is-a-DR-alg},
and we are done. Therefore, the only case we must consider
is when there is a solitary floating vertex $d \in L_i \cup R_i$. Let $t'$ be the first iteration that
started with $L_i \cup R_i$ having a sole vertex $d$ with $x^{t'}(d) \not\in \Z$. 
Then by expanding
\begin{align}
    D_i &\le \left|  \sum_{a \in L_i} v_a X(a) - \sum_{b \in R_i} v_b X(b) \right| \\
    &= \left|  \sum_{a \in L_i} v_a X(a) - \sum_{a \in L_i} v_a x^{t'}(a)  + \sum_{b \in R_i} v_b x^{t'}(b) - \sum_{b \in R_i} v_b X(b) \right| \label{eq:lp-sol-sum-to-zero}\\
    &= \left|  \sum_{a \in L_i} v_a (X(a) - x^{t'}(a)) + \sum_{b \in R_i} v_b (x^{t'}(b) - X(b)) \right| \\
    &\le 
    \sum_{a \in (L_i \cup R_i) - \{d\}} v_a \left| X(a) - x^{t'}(a) \right| 
    + v_d \left| x^{t'}(d) - X(d) \right| \label{eq:Di-apply-tri-ineq} \\
    &< v_d \le v_i^*, \label{eq:vd-bound}
\end{align}
which is our desired $D_i$ bound for \cref{thm:bdr}.
Equation~\eqref{eq:lp-sol-sum-to-zero} follows because we assume
$D_i^1=0$ (as $D_i^0$ corresponds to the LP solution and the pre-processing step thus guarantees $D_i^1=0$) and $t'$ is the
first iteration where $L_i \cup R_i$ contains exactly one floating vertex; therefore, 
by induction and \cref{lemma:Di-preserved}, $D_i^{t'}=0$.
Inequality~\eqref{eq:Di-apply-tri-ineq} follows from the triangle inequality.
The strict inequality in \eqref{eq:vd-bound} follows because $d$ was the sole floating vertex of $L_i \cup R_i$ in iteration $t'$; hence by \cref{lemma:invar-degree}, $\forall a \in (L_i \cup R_i) - \{d\},\; X(a) - x^{t'}(a)=0$ and $|x^{t'}(d) - X(d)| < 1$.

By assumption, $D_i^1=0$ for all $i$
since $\{x_e^0\}_{e \in E}$ is an optimal solution to the corresponding \BSVLP
and the pre-processing step ensures $D_i^0 = 0 \implies D_i^1=0$.
Then, by \cref{lemma:Di-preserved}, the only $D_i$'s
that are not necessarily preserved are those where $L_i \cup R_i$ ends up with exactly one floating vertex in some
algorithm iteration $t'$. As argued above, this case leads to $D_i < v_i^*$. Together with
\cref{lemma:bdr-opt-obj-and-EDi} this completes the proof.
\end{proof}

Consequently from \cref{thm:bdr,thm:bdr-is-a-DR-alg}: 
\begin{corollary}
\BSVLP with all items having equal values is integral.
\end{corollary}
To see this note that if all item values are equal then setting them all equal to $1$ gives an equivalent \BSV instance with $v^*=1$. Then the $D_i < v^*$ guarantee recovers $D_i=0$. In this special case, our randomized algorithm recovers an integral solution to the LP, proving its integrality.

\section{Extension of \BSV to multiple copies of each item}
In \cref{sec:prob-form} we formulated \BSV where each agent has and wants at most one copy of any item $j \in \mathcal{I}$. We now consider \BSVcaps the natural generalization where agents may swap an arbitrary number item copies. 
In addition to the previous inputs, for each agent $i \in \mathcal{N}$, we take capacity functions $\eta_i: \mathcal{I} \to \Z^+$ and $\omega_i: \mathcal{I} \to \Z^+$ denoting the (non-negative) number of copies owned and desired of each $j \in \mathcal{I}$. For simplicity we define $\eta(j)=0$ for $j \not\in H_i$ and $\omega_i(j)=0$ for $j \not\in W_i$.
Analogously, now $g: \mathcal{N} \times \mathcal{N} \times \mathcal{I} \to \Z^+$ and $g(i, i', j) = k$ says \emph{agent $i$ gives $k$ copies of item $j$ to agent $i'$}.
Much like before, a valid allocation is one where each agent $i \in \cN$  (i) receives only desired items \emph{within capacity}, meaning $\sum_{i'\in \cN}g(i',i,j) \leq \omega_i(j)$; and (ii) gives away only items they possess \emph{within capacity}, thus $\sum_{i' \in \cN}g(i,i',j) \leq \eta_i(j)$. 
The goal of \BSVcaps remains the same: to find a valid allocation of maximum utility (now defined as $\sum_{i \in \mathcal{N}} \sum{i' \in \mathcal{N}} \sum_{j \in \mathcal{I}} w(i, i', j) g(i, i', j)$) \emph{subject to} no agent giving away more value than they receive (where the values given and received by agent $i$ are now $\sum{i' \in \mathcal{N}} \sum_{j \in \mathcal{I}} v_j g(i, i', j)$ and $\sum{i' \in \mathcal{N}} \sum_{j \in \mathcal{I}} v_j g(i', i, j)$, respectively). 

Like before, \BSVcaps admits the following similar IP formulation with LP relaxation \BSVLPcap after relaxing \eqref{eq:full-integrality-constr}. For $e = (\ell_{i,j}, r_{i', j})$ we relax $y_e$ to be in $[0, \min \{\eta_i(j), \omega_{i'}(j) \}]$.
\begin{subequations} \label{eq:full-barter-IP}
\begin{align}
    \max \quad            & \sum_{e \in E} w_e x_e \label{eq:full-LP-dynamic}                    \\
    \text{subj. to} \quad & x(\ell_{ij}) \le \eta_i(j), & i\in [n], \ell_{ij} \in L_i&
    \label{eq:full-elli-caps}\\
                          & x(r_{ij}) \le \omega_i(j), & i \in [n], r_{ij} \in R_i&
                          \label{eq:full-ri-caps}\\
                          & \sum_{a \in L_i} x(a) v_a = \sum_{b \in R_i} x(b) v_b, & i \in [n]& \label{eq:full-values-barter} \\
                          & x_e \in \Z^+,
                          & e = (\ell_{ij}, r_{i'j})\in E&.
                          \label{eq:full-integrality-constr}
\end{align}
\end{subequations}
\subsection{Equivalence of \BSV-Caps with single item copies}
By explicitly making \emph{distinct} items for each capacitated item in a \BSVcaps instance we can reduce said instance
to one of \BSV. Although, the resulting \BSV and \VBM formulation now has a number of vertices exponential in the size of the input, this is not a problem. We only use this \BSV instance as a thought experiment to establish that \BDR will
give correct output. To avoid solving any large instances we first solve the polynomial-sized \BSVLPcap and subsequently
observe \BDR actually has polynomial size input in \cref{lemma:at-most-E-floating-vars}.
\begin{lemma} \label{lemma:unit-capacities-suffice}
  Any instance of \BSVcaps has a corresponding (larger but) equivalent \BSV instance with unit capacities.
\end{lemma}%

\begin{proof}[Proof of \cref{lemma:unit-capacities-suffice}]
The instance with unit copies of each item will be large but it is only a thought experiment; we do not
directly solve the corresponding LP nor write the full graph down.
Fix an agent $i$ and an item $j$ in $H_i$.
Make $\eta_i(j)$ copies of this vertex each with unit capacity, say $\ell_{ij1}, \ell_{ij2}, \dots, \ell_{ij\eta_i(j)}$. Similarly, for an item $j' \in W_i$, make $\omega_i(j')$ copies $r_{ij'1}, r_{ij'2}, \dots, r_{ij'\omega_i(j')}$. Like before add edges between all vertices corresponding to the same \emph{original} items. Keep all edge weights the same and use the same corresponding weights for edges between copies i.e., if $e=(\ell_{ijk_1}, r_{i'jk_2})$ and $f=(\ell_{ij}, \ell_{i'j})$ then
$w_e = w_f$. Call this new set of edges over vertex copies $E'$.

To see the two formulations are equivalent, we show $x$, $\forall e \in E, x_e \ge 0$
is feasible to \BSVLPcap if and only if $z$, $\forall e \in E', z_e \in [0,1]$ is feasible to \BSVLP.
Moreover, $x$ and $z$ have the same objective value.
Let $e = (\ell_{ij}, r_{i'j})$ then $x_e = k + r$ for $k \in \Z^+$ and $0 \le r < 1$.
Correspondingly let $e_p = (\ell_{ijp}, r_{i'jp}) \in E'$ and set $z_{e_1}, z_{e_2}, \dots, z_{e_k}$ all equal to $1$
and $z_{e_{k+1}}=r$.
$k+r \le \min \{\eta_i(j), \omega_i(j)\}$ if and only if $x$ and $z$ are feasible.
Moreover, both $x_e$ and $(z_{e_1}, \dots, z_{e_{k+1}})$ each contribute $(k+r)w_e$ to the objective
and $(k+r)v_j$ value given by agent $i$ and received by agent $i'$.

Therefore we always \emph{write and solve} \BSVLPcap and use \BSVLP only as a \emph{thought experiment} to facilitate the presentation of the problem.
It is straightforward to check that the size of \BSVLPcap is $\poly ( |\mathcal{I}|, n, \log \eta, \log \omega)$ where  
$\eta = \max_{i\in [n],j \in H_i} \eta_i(j)$ and similarly $\omega = \max_{i\in [n],j \in H_i} \omega_i(j)$.
\end{proof}
Let $E'$ correspond to the graph with vertex copies as outlined in the proof of \cref{lemma:unit-capacities-suffice}.
\begin{lemma} \label{lemma:at-most-E-floating-vars}
A solution $\{y_e\}_{e \in E'}$ to \BSVLP has at most $|E|$ floating variables.
\end{lemma}
\begin{proof}[Proof of \cref{lemma:at-most-E-floating-vars}]
Like in the proof of \cref{lemma:unit-capacities-suffice}, corresponding to each group of 
$z_{e_1}, z_{e_2}, \dots$ there is at most one $z_{e_p}=r$ for $0 < r < 1$.
Therefore, the number of floating edges is at most $|E|$.
\end{proof}
In conclusion, by \cref{lemma:unit-capacities-suffice,lemma:at-most-E-floating-vars}, \BDR is also an efficient algorithm for the more general \BSVcaps.

\section{Fairness}
Fairness is an important consideration when resource allocation algorithms are deployed in the real-world. \cref{thm:bdr-is-a-DR-alg} allows for adding \textit{fairness} constraints to \BSVLP. Previous works such as \cite{duppala2023GroupFairness,esmaeiliFairLabeledClustering2022,esmaeili2023rawlsian} studied various group fairness notions, and formulated the fair variants of problems like Clustering, Set Packing, etc., by adding \textit{fairness} constraints to the Linear Programs of the respective optimization problems. 

Consider a toy example of such an approach where we are given $\ell$ communities $G_1, \dots, G_\ell \subseteq [n]$ of agents coming together to thicken the market. In order to incentivize said communities to join the centralized exchange, the algorithm designer may promise that each community $G_p$
will receive at least $\mu_p$ units of value \textit{on average}. 
By adding the constraints
\begin{equation}
    \sum_{i \in G_p} x(r_{ij}) v_j \ge \mu_p, \quad   p \in [\ell]
\end{equation}
to the \BSVLP, the algorithm designer ensures that the expected utility of each group $G_p$ is at least $\mu_p$. More precisely, \ref{prop:marginal} and the linearity of expectation ensures
\begin{align*}
\E[\sum_{i \in G_p} X(r_{ij}) v_j] &=  \sum_{i \in G_p} \E[ X(r_{ij})] v_j \\
&= \sum_{i \in G_p} x(r_{ij}) v_j \ge \mu_p.
\end{align*}
The same rationale can be extended to provide individual guarantees (in expectation) by adding analogous constraints for each agent. We conclude this brief discussion by highlighting the versatility of LPs and, as a result, of \BDR.

\section{Hardness of \texorpdfstring{\BSV}{BarterSV}}
We first prove \cref{thm:bsv-hardness}: it is NP-Hard to find any non-empty allocation satisfying $D_k = 0$ for all agents $k$. By non-empty we mean the corresponding LP solution $x \ne 0$, i.e., at least one agent gives away an item. The proof proceeds by reducing from the NP-hard problem of \Ption.
\begin{definition}[\Ption]
    A \Ption instance takes a set $S = \{a_1,a_2,\dots,a_{n} \} $ of $n$ positive integers summing to an integer $2T$. The goal of \Ption is to determine if $S$ can be partitioned into disjoint subsets $S_1$ and $S_2$ such that each subset sums exactly to an integer $T$. 
\end{definition}
\begin{lemma} \label{lemma:hardness-instance-reduction}
    Given a \Ption instance, it can be reduced in polynomial time to a corresponding \BSV instance with two agents. 
\end{lemma}
\begin{proof}
Consider an instance $I = (S,2T)$ of partition problem where $S = \{ a_1,a_2,\dots, a_{n} \}$ is a set of integers where $\sum_{i \in [n]}a_i = 2T$.  
Given $I$, the corresponding \BSV instance is constructed as follows.

Let the set of items $ \mathcal{I} = \{ i_1,i_2,\dots, i_{n}, i_{n+1} \} $ with item values $v_{j} := a_j $ for each item  $i_j, j \in [n]$ and $v_{n+1} := T$ for item $i_{n+1}$. There are only two agents. Agent $1$ has item lists $H_1:=\{i_1,\dots,i_{n}\}$ and $W_1 :=\{i_{n+1}\}$. Symmetrically, agent $2$ has item lists $W_2:=\{ i_{j}: j \in [n]\}$ and $H_2:=\{i_{n+1}\}$. 
The particular weights $w_e$ of allocating items (i.e., allocation's utility) do not matter as we only care about whether some non-empty allocation exists. 
\end{proof}
Recall the goal is to show there exists a \emph{non-empty} allocation 
such that for each agent $k\in [2]$, $D_k = 0$ if and only if the corresponding \Ption instance has a solution.
\begin{lemma}\label{lemma:hardness1}
    There exists a polynomial time algorithm to find a non-empty allocation of items with $D_k=0$ for each agent $k$ in the \BSV instance if and only if there exists a polynomial time algorithm to the corresponding \Ption instance. 
\end{lemma}
\begin{proof}
\textbf{Forward direction} (\Ption $\implies$ \BSV ).
Given a solution $(S_1,S_2)$ to the \Ption instance, the corresponding \BSV instance has a solution in the following manner. Allocate the items $\{ i_j : j \in S_1 \} \subseteq H_1$ to agent $2$ and allocate the item $i_{n+1} \in H_2 $ to agent $1$. Thus, the value of the items received and given by both the agents is exactly $T$ resulting in a non-empty allocation with $D_{1}= D_2=0$. 

\textbf{Backward Direction} (\BSV $\implies$ \Ption).
Take a \emph{non-empty} allocation of items $\mathcal{I}$ to each agent $k \in [2]$ with $D_k =0$.
Such a non-empty allocation must have agent 
$1$ giving away their only item, which has value $T$.
Therefore $D_1=0$ implies agent $1$ received $T$ units of value. 
Let the items agent $1$ received be $i_{j_1}, i_{j_2}, \dots, i_{j_\ell}$, letting $J = \{ j_1, \dots, j_\ell\}$.
Thus, $\sum_{p \in J} v_{p} = T$.
Therefore, the corresponding partition instance has solution $S_1 = \{ a_p : p \in J \}$ and $S_2 = \{ a_p : p \not\in J\}$.
\end{proof}

Together, \cref{lemma:hardness-instance-reduction,lemma:hardness1} prove \cref{thm:bsv-hardness}: it is NP-hard to find a non-empty allocation of items with $D_k=0$ for any agent $k$.

\subsection{LP approaches will not beat \texorpdfstring{$D_i < v^*$}{Di < v*}}
A natural follow-up question: is it possible to guarantee $D_i < v^* - \epsilon$ for some
$\epsilon > 0 $? At least using \BSVLP, this is not be possible. Consider the family of \BSV instances
$\Pi(n)$ wherein two agents have one item each and desire each other's item. The
agent's items are $j_1$ and $j_2$, respectively, with values $v_{j_1}=1$ and $v_{j_2}=1/n$. Let the utility be that of maximizing the total value reallocated (i.e., for $e=(a,b)$, $w_e=v_a=v_b$, $\forall e \in E$ in the corresponding VBM instance).
The LP solution corresponds to agent 1 giving a $1/n$-fraction of her item away and agent 2 giving his entire item away for an optimal objective value of $2/n$. 
Even if agent $2$ always gives his item away, the objective achieved is only $1/n$.
Therefore, to achieve the optimal objective agent $1$ must give away her item with positive probability.
This leads to the worst case, $D_1 \ge 1 - 1 / n = v^* - 1/n$. Therefore,
\begin{remark}
    For any $\epsilon > 0$, there exists $n$ such that $\Pi(n)$ achieves $D_i \ge v^* - \epsilon$.
\end{remark}

A moment's reflection makes it clear that the only feasible (integral) reallocation to this family of instances is the empty reallocation i.e., the zero vector. A direction to bypass this "integrality gap"-like barrier in $D_i$ might be to write \BSVLP only for instances in which we know $D_i=0$ has a feasible non-empty integral reallocation. However, this is in and of itself a hard problem as proved in \cref{thm:bsv-hardness}.

\section{Conclusion}
We introduce and study \BSV, a centralized barter exchange problem where each item has a value agreed upon by the participating agents. The goal is to find an allocation/exchange that (i) maximizes the collective utility of the allocation such that (ii) the total value of each agent's items before and after the exchange is equal.
Though it is NP-hard to solve \BSV exactly, we can efficiently compute allocations with optimal expected utility where each agent's \textit{net value loss} is at most a single item's value. Our problem is motivated by the proliferation of large scale web markets on social media websites with 50,000-60,000 active users eager to swap items with one another. We formulate and study this novel problem with several real-world exchanges of video games, board games, digital goods and more. These exchanges have large communities, but their decentralized nature leaves much to be desired in terms of efficiency.

Future directions of this work include accounting for arbitrary item valuations i.e., different agents may value items differently. Directly applying our algorithm does not work because we rely on the paths of $G$ consisting of items with all-equal values. Additionally, we briefly touch on how some fairness criteria interact with our model algorithm; however, the exploration of many important fairness criteria is left for future work. Finally, we argue why our specific LP-based approach cannot guarantee better than $D_i < v^*$. It remains an open question whether other approaches (LP-based or not) can beat this bound or whether improving this result is NP-hard altogether.

\begin{acks}
    Sharmila Duppala, Juan Luque and Aravind Srinivasan are all supported in part by NSF grant CCF-1918749.
\end{acks}

\bibliographystyle{ACM-Reference-Format}
\bibliography{refs}


\begin{thebibliography}{23}


\ifx \showCODEN    \undefined \def \showCODEN     #1{\unskip}     \fi
\ifx \showDOI      \undefined \def \showDOI       #1{#1}\fi
\ifx \showISBNx    \undefined \def \showISBNx     #1{\unskip}     \fi
\ifx \showISBNxiii \undefined \def \showISBNxiii  #1{\unskip}     \fi
\ifx \showISSN     \undefined \def \showISSN      #1{\unskip}     \fi
\ifx \showLCCN     \undefined \def \showLCCN      #1{\unskip}     \fi
\ifx \shownote     \undefined \def \shownote      #1{#1}          \fi
\ifx \showarticletitle \undefined \def \showarticletitle #1{#1}   \fi
\ifx \showURL      \undefined \def \showURL       {\relax}        \fi
\providecommand\bibfield[2]{#2}
\providecommand\bibinfo[2]{#2}
\providecommand\natexlab[1]{#1}
\providecommand\showeprint[2][]{arXiv:#2}

\bibitem[Abbassi et~al\mbox{.}(2013)]%
        {abbassi2013fair}
\bibfield{author}{\bibinfo{person}{Zeinab Abbassi}, \bibinfo{person}{Laks V.~S.
  Lakshmanan}, {and} \bibinfo{person}{Min Xie}.}
  \bibinfo{year}{2013}\natexlab{}.
\newblock \showarticletitle{Fair Recommendations for Online Barter Exchange
  Networks}. In \bibinfo{booktitle}{\emph{International Workshop on the Web and
  Databases}}.
\newblock


\bibitem[Abraham et~al\mbox{.}(2007)]%
        {abrahamClearingAlgorithmsBarter2007}
\bibfield{author}{\bibinfo{person}{David~J. Abraham}, \bibinfo{person}{Avrim
  Blum}, {and} \bibinfo{person}{Tuomas Sandholm}.}
  \bibinfo{year}{2007}\natexlab{}.
\newblock \showarticletitle{Clearing Algorithms for Barter Exchange Markets:
  Enabling Nationwide Kidney Exchanges}. In
  \bibinfo{booktitle}{\emph{Proceedings of the 8th {{ACM}} Conference on
  {{Electronic}} Commerce}} \emph{(\bibinfo{series}{{{EC}} '07})}.
  \bibinfo{publisher}{{Association for Computing Machinery}},
  \bibinfo{address}{{New York, NY, USA}}, \bibinfo{pages}{295--304}.
\newblock
\showISBNx{978-1-59593-653-0}
\urldef\tempurl%
\url{https://doi.org/10.1145/1250910.1250954}
\showDOI{\tempurl}


\bibitem[Ashlagi et~al\mbox{.}(2018)]%
        {ashlagiEffectMatchrunFrequencies2018}
\bibfield{author}{\bibinfo{person}{Itai Ashlagi}, \bibinfo{person}{Adam
  Bingaman}, \bibinfo{person}{Maximilien Burq}, \bibinfo{person}{Vahideh
  Manshadi}, \bibinfo{person}{David Gamarnik}, \bibinfo{person}{Cathi Murphey},
  \bibinfo{person}{Alvin~E. Roth}, \bibinfo{person}{Marc~L. Melcher}, {and}
  \bibinfo{person}{Michael~A. Rees}.} \bibinfo{year}{2018}\natexlab{}.
\newblock \showarticletitle{Effect of Match-Run Frequencies on the Number of
  Transplants and Waiting Times in Kidney Exchange}.
\newblock \bibinfo{journal}{\emph{American Journal of Transplantation}}
  \bibinfo{volume}{18}, \bibinfo{number}{5} (\bibinfo{year}{2018}),
  \bibinfo{pages}{1177--1186}.
\newblock
\showISSN{1600-6143}
\urldef\tempurl%
\url{https://doi.org/10.1111/ajt.14566}
\showDOI{\tempurl}


\bibitem[Aziz et~al\mbox{.}(2021)]%
        {azizFairAllocationIndivisible2021}
\bibfield{author}{\bibinfo{person}{Haris Aziz}, \bibinfo{person}{Ioannis
  Caragiannis}, \bibinfo{person}{Ayumi Igarashi}, {and} \bibinfo{person}{Toby
  Walsh}.} \bibinfo{year}{2021}\natexlab{}.
\newblock \showarticletitle{Fair Allocation of Indivisible Goods and Chores}.
\newblock \bibinfo{journal}{\emph{Auton Agent Multi-Agent Syst}}
  \bibinfo{volume}{36}, \bibinfo{number}{1} (\bibinfo{date}{Nov.}
  \bibinfo{year}{2021}), \bibinfo{pages}{3}.
\newblock
\showISSN{1573-7454}
\urldef\tempurl%
\url{https://doi.org/10.1007/s10458-021-09532-8}
\showDOI{\tempurl}


\bibitem[Bir{\'o} et~al\mbox{.}(2009)]%
        {biroMaximumWeightCycle2009}
\bibfield{author}{\bibinfo{person}{P{\'e}ter Bir{\'o}},
  \bibinfo{person}{David~F. Manlove}, {and} \bibinfo{person}{Romeo Rizzi}.}
  \bibinfo{year}{2009}\natexlab{}.
\newblock \showarticletitle{Maximum Weight Cycle Packing in Directed Graphs,
  with Application to Kidney Exchange Programs}.
\newblock \bibinfo{journal}{\emph{Discrete Math. Algorithm. Appl.}}
  \bibinfo{volume}{01}, \bibinfo{number}{04} (\bibinfo{date}{Dec.}
  \bibinfo{year}{2009}), \bibinfo{pages}{499--517}.
\newblock
\showISSN{1793-8309}
\urldef\tempurl%
\url{https://doi.org/10.1142/S1793830909000373}
\showDOI{\tempurl}


\bibitem[Brubach et~al\mbox{.}(2017)]%
        {Brubach2017AlgorithmsTA}
\bibfield{author}{\bibinfo{person}{Brian Brubach},
  \bibinfo{person}{Karthik~Abinav Sankararaman}, \bibinfo{person}{Aravind
  Srinivasan}, {and} \bibinfo{person}{Pan Xu}.}
  \bibinfo{year}{2017}\natexlab{}.
\newblock \showarticletitle{Algorithms to Approximate Column-sparse Packing
  Problems}.
\newblock \bibinfo{journal}{\emph{ACM Transactions on Algorithms (TALG)}}
  \bibinfo{volume}{16} (\bibinfo{year}{2017}), \bibinfo{pages}{1 -- 32}.
\newblock


\bibitem[Carvalho et~al\mbox{.}(2021)]%
        {carvalho2021robust}
\bibfield{author}{\bibinfo{person}{Margarida Carvalho}, \bibinfo{person}{Xenia
  Klimentova}, \bibinfo{person}{Kristiaan Glorie}, \bibinfo{person}{Ana Viana},
  {and} \bibinfo{person}{Miguel Constantino}.} \bibinfo{year}{2021}\natexlab{}.
\newblock \showarticletitle{Robust models for the kidney exchange problem}.
\newblock \bibinfo{journal}{\emph{INFORMS Journal on Computing}}
  \bibinfo{volume}{33}, \bibinfo{number}{3} (\bibinfo{year}{2021}),
  \bibinfo{pages}{861--881}.
\newblock


\bibitem[Chekuri et~al\mbox{.}(2010)]%
        {chekuri2010dependent}
\bibfield{author}{\bibinfo{person}{Chandra Chekuri}, \bibinfo{person}{Jan
  Vondr{\'a}k}, {and} \bibinfo{person}{Rico Zenklusen}.}
  \bibinfo{year}{2010}\natexlab{}.
\newblock \showarticletitle{Dependent randomized rounding via exchange
  properties of combinatorial structures}. In \bibinfo{booktitle}{\emph{2010
  IEEE 51st Annual Symposium on Foundations of Computer Science}}. IEEE,
  \bibinfo{pages}{575--584}.
\newblock


\bibitem[Constantino et~al\mbox{.}(2013)]%
        {constantino2013new}
\bibfield{author}{\bibinfo{person}{Miguel Constantino}, \bibinfo{person}{Xenia
  Klimentova}, \bibinfo{person}{Ana Viana}, {and} \bibinfo{person}{Abdur
  Rais}.} \bibinfo{year}{2013}\natexlab{}.
\newblock \showarticletitle{New insights on integer-programming models for the
  kidney exchange problem}.
\newblock \bibinfo{journal}{\emph{European Journal of Operational Research}}
  \bibinfo{volume}{231}, \bibinfo{number}{1} (\bibinfo{year}{2013}),
  \bibinfo{pages}{57--68}.
\newblock


\bibitem[Duppala et~al\mbox{.}(2023)]%
        {duppala2023GroupFairness}
\bibfield{author}{\bibinfo{person}{Sharmila Duppala}, \bibinfo{person}{Juan
  Luque}, \bibinfo{person}{John Dickerson}, {and} \bibinfo{person}{Aravind
  Srinivasan}.} \bibinfo{year}{2023}\natexlab{}.
\newblock \showarticletitle{Group Fairness in Set Packing Problems}. In
  \bibinfo{booktitle}{\emph{Proceedings of the Thirty-Second International
  Joint Conference on Artificial Intelligence, {IJCAI-23}}},
  \bibfield{editor}{\bibinfo{person}{Edith Elkind}} (Ed.).
  \bibinfo{publisher}{International Joint Conferences on Artificial
  Intelligence Organization}, \bibinfo{pages}{391--399}.
\newblock
\urldef\tempurl%
\url{https://doi.org/10.24963/ijcai.2023/44}
\showDOI{\tempurl}
\newblock
\shownote{Main Track}.


\bibitem[Esmaeili et~al\mbox{.}(2023)]%
        {esmaeili2023rawlsian}
\bibfield{author}{\bibinfo{person}{Seyed Esmaeili}, \bibinfo{person}{Sharmila
  Duppala}, \bibinfo{person}{Davidson Cheng}, \bibinfo{person}{Vedant Nanda},
  \bibinfo{person}{Aravind Srinivasan}, {and} \bibinfo{person}{John~P
  Dickerson}.} \bibinfo{year}{2023}\natexlab{}.
\newblock \showarticletitle{Rawlsian fairness in online bipartite matching:
  Two-sided, group, and individual}. In \bibinfo{booktitle}{\emph{Proceedings
  of the AAAI Conference on Artificial Intelligence}},
  Vol.~\bibinfo{volume}{37}. \bibinfo{pages}{5624--5632}.
\newblock


\bibitem[Esmaeili et~al\mbox{.}(2022)]%
        {esmaeiliFairLabeledClustering2022}
\bibfield{author}{\bibinfo{person}{Seyed~A. Esmaeili},
  \bibinfo{person}{Sharmila Duppala}, \bibinfo{person}{John~P. Dickerson},
  {and} \bibinfo{person}{Brian Brubach}.} \bibinfo{year}{2022}\natexlab{}.
\newblock \bibinfo{title}{Fair {{Labeled Clustering}}}.
\newblock
\newblock
\urldef\tempurl%
\url{https://doi.org/10.48550/arXiv.2205.14358}
\showDOI{\tempurl}
\showeprint[arxiv]{2205.14358}~[cs]


\bibitem[Fang et~al\mbox{.}(2015)]%
        {fang2015randomized}
\bibfield{author}{\bibinfo{person}{Wenyi Fang}, \bibinfo{person}{Aris
  Filos-Ratsikas}, \bibinfo{person}{S{\o}ren Kristoffer~Stiil Frederiksen},
  \bibinfo{person}{Pingzhong Tang}, {and} \bibinfo{person}{Song Zuo}.}
  \bibinfo{year}{2015}\natexlab{}.
\newblock \showarticletitle{Randomized assignments for barter exchanges:
  Fairness vs. efficiency}. In \bibinfo{booktitle}{\emph{Algorithmic Decision
  Theory: 4th International Conference, ADT 2015, Lexington, KY, USA, September
  27--30, 2015, Proceedings}}. Springer, \bibinfo{pages}{537--552}.
\newblock


\bibitem[Farnadi et~al\mbox{.}(2021)]%
        {farnadiIndividualFairnessKidney2021}
\bibfield{author}{\bibinfo{person}{Golnoosh Farnadi}, \bibinfo{person}{William
  {St-Arnaud}}, \bibinfo{person}{Behrouz Babaki}, {and}
  \bibinfo{person}{Margarida Carvalho}.} \bibinfo{year}{2021}\natexlab{}.
\newblock \showarticletitle{Individual {{Fairness}} in {{Kidney Exchange
  Programs}}}.
\newblock \bibinfo{journal}{\emph{Proceedings of the AAAI Conference on
  Artificial Intelligence}} \bibinfo{volume}{35}, \bibinfo{number}{13}
  (\bibinfo{date}{May} \bibinfo{year}{2021}), \bibinfo{pages}{11496--11505}.
\newblock
\showISSN{2374-3468}


\bibitem[Gandhi et~al\mbox{.}(2006)]%
        {gandhiDependentRoundingIts2006}
\bibfield{author}{\bibinfo{person}{Rajiv Gandhi}, \bibinfo{person}{Samir
  Khuller}, \bibinfo{person}{Srinivasan Parthasarathy}, {and}
  \bibinfo{person}{Aravind Srinivasan}.} \bibinfo{year}{2006}\natexlab{}.
\newblock \showarticletitle{Dependent Rounding and Its Applications to
  Approximation Algorithms}.
\newblock \bibinfo{journal}{\emph{J. ACM}} \bibinfo{volume}{53},
  \bibinfo{number}{3} (\bibinfo{date}{May} \bibinfo{year}{2006}),
  \bibinfo{pages}{324--360}.
\newblock
\showISSN{0004-5411}
\urldef\tempurl%
\url{https://doi.org/10.1145/1147954.1147956}
\showDOI{\tempurl}


\bibitem[Glorie et~al\mbox{.}(2014)]%
        {glorie2014kidney}
\bibfield{author}{\bibinfo{person}{Kristiaan~M Glorie},
  \bibinfo{person}{J~Joris van~de Klundert}, {and} \bibinfo{person}{Albert~PM
  Wagelmans}.} \bibinfo{year}{2014}\natexlab{}.
\newblock \showarticletitle{Kidney exchange with long chains: An efficient
  pricing algorithm for clearing barter exchanges with branch-and-price}.
\newblock \bibinfo{journal}{\emph{Manufacturing \& Service Operations
  Management}} \bibinfo{volume}{16}, \bibinfo{number}{4}
  (\bibinfo{year}{2014}), \bibinfo{pages}{498--512}.
\newblock


\bibitem[Herlihy(2018)]%
        {Herlihy18:Atomic}
\bibfield{author}{\bibinfo{person}{Maurice Herlihy}.}
  \bibinfo{year}{2018}\natexlab{}.
\newblock \showarticletitle{Atomic cross-chain swaps}. In
  \bibinfo{booktitle}{\emph{ACM Symposium on Principles of Distributed
  Computing (PODC)}}. \bibinfo{pages}{245--254}.
\newblock


\bibitem[Jevons(1879)]%
        {Jevons79:Theory}
\bibfield{author}{\bibinfo{person}{William~Stanley Jevons}.}
  \bibinfo{year}{1879}\natexlab{}.
\newblock \bibinfo{booktitle}{\emph{The Theory of Political Economy}}.
\newblock \bibinfo{publisher}{Macmillan and Company}.
\newblock


\bibitem[Klimentova et~al\mbox{.}(2021)]%
        {klimentova2021fairness}
\bibfield{author}{\bibinfo{person}{Xenia Klimentova}, \bibinfo{person}{Ana
  Viana}, \bibinfo{person}{Jo{\~a}o~Pedro Pedroso}, {and}
  \bibinfo{person}{Nicolau Santos}.} \bibinfo{year}{2021}\natexlab{}.
\newblock \showarticletitle{Fairness models for multi-agent kidney exchange
  programmes}.
\newblock \bibinfo{journal}{\emph{Omega}}  \bibinfo{volume}{102}
  (\bibinfo{year}{2021}), \bibinfo{pages}{102333}.
\newblock


\bibitem[McElfresh et~al\mbox{.}(2020)]%
        {mcelfreshImprovingPolicyConstrainedKidney2020}
\bibfield{author}{\bibinfo{person}{Duncan McElfresh}, \bibinfo{person}{Michael
  Curry}, \bibinfo{person}{Tuomas Sandholm}, {and} \bibinfo{person}{John
  Dickerson}.} \bibinfo{year}{2020}\natexlab{}.
\newblock \showarticletitle{Improving Policy-Constrained Kidney Exchange via
  Pre-Screening}. In \bibinfo{booktitle}{\emph{Advances in Neural Information
  Processing Systems}}, Vol.~\bibinfo{volume}{33}. \bibinfo{pages}{2674--2685}.
\newblock


\bibitem[Noothigattu et~al\mbox{.}(2020)]%
        {noothigattu2020axioms}
\bibfield{author}{\bibinfo{person}{Ritesh Noothigattu},
  \bibinfo{person}{Dominik Peters}, {and} \bibinfo{person}{Ariel~D Procaccia}.}
  \bibinfo{year}{2020}\natexlab{}.
\newblock \showarticletitle{Axioms for learning from pairwise comparisons}.
\newblock \bibinfo{journal}{\emph{Advances in Neural Information Processing
  Systems}}  \bibinfo{volume}{33} (\bibinfo{year}{2020}),
  \bibinfo{pages}{17745--17754}.
\newblock


\bibitem[Plaut et~al\mbox{.}(2016)]%
        {plaut2016fast}
\bibfield{author}{\bibinfo{person}{Benjamin Plaut}, \bibinfo{person}{John
  Dickerson}, {and} \bibinfo{person}{Tuomas Sandholm}.}
  \bibinfo{year}{2016}\natexlab{}.
\newblock \showarticletitle{Fast optimal clearing of capped-chain barter
  exchanges}. In \bibinfo{booktitle}{\emph{Proceedings of the AAAI Conference
  on Artificial Intelligence}}, Vol.~\bibinfo{volume}{30}.
\newblock


\bibitem[Thyagarajan et~al\mbox{.}(2022)]%
        {Thyagarajan22:Universal}
\bibfield{author}{\bibinfo{person}{Sri~AravindaKrishnan Thyagarajan},
  \bibinfo{person}{Giulio Malavolta}, {and} \bibinfo{person}{Pedro
  Moreno-Sanchez}.} \bibinfo{year}{2022}\natexlab{}.
\newblock \showarticletitle{Universal atomic swaps: Secure exchange of coins
  across all blockchains}. In \bibinfo{booktitle}{\emph{IEEE Symposium on
  Security and Privacy (S\&P)}}. IEEE, \bibinfo{pages}{1299--1316}.
\newblock


\end{thebibliography}

\appendix
\section{Direct application of \GKPSDR fails}
\begin{figure}[h!]
    \centering
    \includegraphics[scale = 0.5]{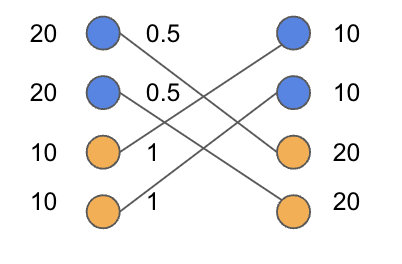}
\caption{The bipartite graph corresponding to the \BSV instance in \cref{ex:worstcase}is pictured. Blue and orange vertices correspond to agents $1$ and $2$, respectively. The optimal LP solution is $x = [ 0.5,0.5,1,1]$. 
    \label{fig:worstcase}
}
\end{figure}

\cref{ex:worstcase} is a worst case instance where a direct application of \GKPSDR to the fractional optimal solution of \BSV results in a net loss of $\sum_{j \in L_i}v_j$ for some  agent $i$; i.e., agent $i$ gives away all their items and does not receive any item from its wishlist. 
\begin{example}\label{ex:worstcase}
    Consider an instance of \BSV with two agents where $W_1 = \{1,2\},W_2 = \{3,4\}$ and $H_1 = \{3,4\}, H_2= \{1,2\}$. Let the values of the items be $v(1)=v(2)=10$ and $v(3)=v(4)=20$. 
\end{example}
\cref{fig:worstcase} shows the bipartite graph of the \BSV instance from \cref{ex:worstcase}. The edges are unweighted and the optimal LP solution is $x = [0.5,0.5,1,1]$. The first two coordinates of $x$ correspond to items given by agent $1$. \GKPSDR will round both of these co-ordinates to $0$ with positive probability thus resulting in agent $2$ incurring a net loss of $\sum_{a \in L_2} v_a = 20$ units of value.    

\begin{observation}
    \GKPSDR rounding $x$ results in the vector $X= [0,0,1,1]$ with positive probability; this is the worst case for agent $2$ where their net loss is $\sum_{a \in L_2}v_a = 20$ units of value.
\end{observation}
The above example can be easily generalized to instances with larger item-lists and more agents where some agent $i$ achieves net value loss $\sum_{ a \in L_i}v_a$ with positive probability.

\section{Proof of Theorem~\ref{thm:bdr-is-a-DR-alg}} 
\label{sec:proof-of-bdr-is-a-DR-alg}
\begin{lemma} \label{lemma:invar-marginal}
    \BDR satisfies \ref{invar:marginal}.
\end{lemma}
\begin{proof}
    The property holds trivially for $r=0$. Recall $r=1$ corresponds to the output of the pre-processing step. This fact is proved in \cite{gandhiDependentRoundingIts2006}. Therefore, focus on some fixed $r>1$ and proceed by induction. Fix $e \in E$ and the CCC/CCW to be rounded $\Pc = \mCCC$.
    Proceed by considering the following two events.
    
    \emph{Event $A$:}
    $e$ does not appear in $\Pc$ so $e$ does not change this iteration. Thus by the induction hypothesis $\E[x_e^{r+1} \mid (x_e^r = z) \wedge A] = z$.\\
    \emph{Event $B$:}
    $e$ appears in $\Pc$, say, in path $s_i \leadsto t_i$ for a fixed $i$.
    Recall values $\alpha$ and $\beta$ from \eqref{eq:bdr-alpha-beta} are fixed and
    $x_e^{r}$ is modified according to \eqref{eq:round-event-A}.
    Assuming $e \in M^i_{-1}$ then
    $$\E[x_e^{r+1} \mid (x_e^r = z) \wedge B] = z + \frac{\alpha}{v_i} \left(\frac{\beta}{\alpha + \beta} \right) - \frac{\beta}{v_i} \left( \frac{\alpha}{\alpha + \beta}\right) = z.$$
    The same holds if instead $e \in M^i_{1}$. Hence
    \begin{align*}
    \E[x_e^{r+1} \mid (x_e^r=z)] &= 
    \E[x_e^{r+1} \mid (x_e^r=z) \wedge B] \cdot \Pr(B)\\
    &\phantom{=}+ 
    \E[x_e^{r+1} \mid (x_e^r=z) \wedge A] \cdot \Pr(A) \\
    &= z (\Pr(A) + \Pr(B)) = z.
    \end{align*}
    
    Let $Z$ be the set of possible values for $x_e^r$. 
    \begin{align*}
        \E[x_e^{r+1}] &= 
        \sum_{z \in Z} \E[x_e^{r+1} \mid (x_e^r = z)] \cdot \Pr(x_e^r = z) \\
        &= \sum_{z \in Z} z \cdot \Pr(x_e^r = z) = \E[x_e^r].
    \end{align*}
    By the IH, then $\E[x_e^{r+1}] = x_e^0$.
\end{proof}

\begin{lemma} \label{lemma:invar-degree}
    \BDR satisfies \ref{invar:degree}.
\end{lemma}
\begin{proof}
    The property holds trivially for $r=0$. Recall $r=1$ corresponds to the output of the pre-processing step. This fact is proved in \cite{gandhiDependentRoundingIts2006}. Therefore, focus on some fixed $r>1$ and proceed by induction. Fix $a \in L \cup R$ and the CCC/CCW to be rounded $\Pc = \mCCC$.
    Recall $V(\Pc)$ denotes the endpoints of the paths of $\Pc$.
    Proceed by cases.

    \emph{Case $A$:} $a \not \in V(\Pc)$. Then either $a$ does not appear in $\Pc$
    or $a$ appears in $\Pc$ but with two edges incident on it. In the former case clearly $x_a^{r+1} = x_a^r$.
    In the latter case, the change of each incident edge is equal in magnitude and opposite in sign (since one edge belongs to $M^i_{-1}$ and the other to $M^i_1$) therefore $x_a^{r+1} = x_a^r$ as well.
    Thus by the IH $\lfloor x^0(a) \rfloor\le x^{r+1}(a) \le  \lceil x^0(a) \rceil$.

    \emph{Case $B$:} $a \in V(\Pc)$. There is a single incident edge $e \in N(a)$.
    Without loss of generality, said edge belongs path $s_i \leadsto t_i$ and thus to $M^i_{-1}$ (the proof for $M^i_{1}$ is identical). Then either 
    $x^{r+1}(a) = x^{r}(a) + \alpha / v_i $ or $x^{r+1}(a) = x^r(a) - \beta / v_i$. 
    In either case, by definition of $\alpha$ and $\beta$ (i.e., \eqref{eq:bdr-alpha-beta}), $\alpha$ and $\beta$ are small enough that 
    $\lfloor x^r(a) \rfloor\le x^{r+1}(a) \le  \lceil x^r(a) \rceil$.
    Observe $\lfloor x^0(a) \rfloor = \lfloor x^r(a) \rfloor$
    and $\lceil x^0(a) \rceil = \lceil x^r(a) \rceil$.

    Having handled exhaustive cases, the proof is complete.
\end{proof}

\begin{lemma} \label{lemma:invar-neg-corr}
    \BDR satisfies \ref{invar:neg-corr}.
\end{lemma}
\begin{proof}
    The property holds trivially for $r=0$. Recall $r=1$ corresponds to the output of the pre-processing step. This fact is proved in \cite{gandhiDependentRoundingIts2006}. Therefore, focus on some fixed $r>1$ and proceed by induction.
    Fix a vertex $a$ and a subset of edges $S$ incident on $a$ like in
    \ref{invar:neg-corr}.
    Also fix the CCC/CCW to be rounded $\Pc = \mCCC$. Proceed based on the following events. 

    \emph{Event $A$:} no edge in $S$ has its value modified. Then 
    $\E[\prod_{e \in S} x^{r+1}_e \mid A] =  \E[\prod_{e \in S} x^{r}_e \mid A] $.

    \emph{Event $B$:} two edges $e_1, e_2 \in S$ have their values modified. 
    Said edges must both belong to $s_i \leadsto t_i$, for some fixed $i$, with one belonging to $M^i_{-1}$ and the other to $M^i_{1}$; say $e_1 \in M^i_1$ and $e_2 \in M^i_{-1}$.
    Then 
    $$
    (x_{e_1}^{r+1}, x_{e_2}^{r+1}) = 
    \begin{cases}
        (x_{e_1}^{r} + \alpha/v_i, x_{e_2}^r - \alpha/v_i) 
        \text{ with probability } \beta / (\alpha + \beta) \\ 
        (x_{e_1}^{r} - \beta/v_i, x_{e_2}^r + \beta/v_i) 
        \text{ with probability } \alpha / (\alpha + \beta)
    \end{cases}
    $$
    where $\alpha$ and $\beta$ are fixed per \eqref{eq:bdr-alpha-beta}.
    Let $S_1 = S - \{ e_1, e_2\}$. Then 
    \begin{equation*}
    \E\left[ \prod_{e \in S} x_e^{r+1} \mid (\forall e \in S, x_e^r=z_e) \wedge B \right] 
    = 
    \E\left[ x_{e_1}^r \cdot x_{e_2}^r \mid (\forall e \in S, x_e^r=z_e) \wedge B  \right] \prod_{e \in S_1} z_e.
    \end{equation*}
    The above expectation can be written as $(\Psi + \Phi) \prod_{e \in S_1} z_e$, where
    \begin{align*}
        \Psi &= (\beta / (\alpha + \beta)) \cdot (z_{e_1} + \alpha)  \cdot (z_{e_2} - \alpha) \text{ and }\\
        \Phi &= (\alpha / (\alpha + \beta)) \cdot (z_{e_1} - \beta)  \cdot (z_{e_2} + \beta).
    \end{align*}
    It is easy to s how $\Psi + \Phi \le z_{e_1} z_{e_2}$. Thus, for any fixed $\{e_1, e_2\} \subseteq S$ and for any fixed $(\alpha, \beta)$, and for fixed values of $z_e$, the following holds:
    \begin{equation*}
    \E\left[ \prod_{e \in S} x_e^{r+1} \mid (\forall e \in S, x_e^r=z_e) \wedge B \right]
    \le \prod_{e \in S} z_e.
    \end{equation*}
    Hence, $\E[\prod_{e \in S} x_e^{r+1} \mid B] \le \E[\prod_{e \in S} x_e^{r} \mid B]$. 

    \emph{Event $C$:} exactly one edge in the set $S$ has its value modified. Let $C$
    denote the event that edge $e_1 \in S$ has its value changed in the following
    probabilistic way
    \begin{equation*}
        x_{e_1}^{r+1} =
        \begin{cases}
            x_{e_1}^r + \alpha \text{ with probability } \beta/(\alpha + \beta) \\
            x_{e_1}^r - \beta \text{ with probability } \alpha/(\alpha + \beta).
        \end{cases}
    \end{equation*}
    Thus, $\E[ x_{e_1}^{r+1} \mid (\forall e \in S, x_{e}^{r} = z_e) \wedge C] = z_{e_1}$. Letting $S_1 = S - \{e_1\}$, we get that
    $\E[ \prod_{e \in S} x_e^{r+1} \mid (\forall e \in S, x_e^{r} = z_f) \wedge C ]$ equals
    \begin{equation*}
    E[ x_{e_1}^{r+1} \mid (\forall e \in S, x_e^{r} = z_f) \wedge C ]
    \prod_{e \in S_1} z_e = \prod_{e \in S} z_e.
    \end{equation*}
    Since the equation holds for any $e_1 \in S$, for any values of $z_e$, and for any $(\alpha, \beta)$, we have
    $\E[\prod_{e \in S} x_e^{r+1} \mid C] = \E[\prod_{e \in S} x_e^r]$.
\end{proof}

\begin{proof}[Proof of \cref{thm:bdr-is-a-DR-alg}]
By \cref{lemma:invar-marginal,lemma:invar-degree} \BDR satisfies \ref{prop:marginal} and \ref{prop:degree}. Let $T$ be the last iteration of \BDR. From \cref{lemma:invar-neg-corr} we have
\begin{equation*}
\Pr(\bigwedge_{e \in S} (X_e = 1)) = \E[\prod_{e \in S} x_e^{T+1}] \le
\E[\prod_{e \in S} x_e^{1}] = \prod_{e \in S} x_e^0 = \prod_{e \in S} \Pr(X_e = 1).
\end{equation*}
The proof for $c=0$ (i.e., $\Pr(X_e = 0)$) is identical. Therefore, \BDR satisfies
\ref{prop:neg-corr}.
\end{proof}

\end{document}